\documentclass[ijoc,nonblindrev]{informs3} 

\OneAndAHalfSpacedXII 

\usepackage{amsfonts,amssymb,amsbsy,mathrsfs} 
\usepackage{bm}
\usepackage{algorithm,algorithmic}
\usepackage{accents}
\usepackage{dsfont}
\usepackage{epstopdf}
\usepackage{float}
\usepackage{subfig}
\usepackage{lineno}
\usepackage{color}
\usepackage{eucal}
\usepackage{accents}
\usepackage{epstopdf}
\usepackage{graphicx}
\usepackage{tikz}
\usepackage[utf8]{inputenc} 

\newcommand{\y}{\bm{y}}
\newcommand{\Y}{\bm{Y}}
\newcommand{\f}{\bm{f}}
\newcommand{\g}{\bm{g}}

\newcommand{\mcA}{\mathscr{A}}

\newcommand{\mcC}{\mathcal{C}}

\newcommand{\mbC}{\mathbb{C}}

\newcommand{\mbR}{\mathbb{R}}

\newcommand{\xv}{\bm{x}}

\newcommand{\intC}{\mbC^{\circ}}
 
\newcommand{\ind}[1]{\mathds{1}\left\{#1\right\}} 
\newcommand{\inD}{\Rightarrow}
\newcommand{\real}{\mbR}
\newcommand{\ti}{t_\text{min}}
\newcommand{\ta}{t_\text{max}}
\newcommand{\refeq}[1]{{(\ref{#1})}}

\usepackage{natbib}
 \bibpunct[, ]{(}{)}{,}{a}{}{,}%
 \def\bibfont{\small}%
\TheoremsNumberedThrough     

\EquationsNumberedThrough    

\begin{document}


\RUNAUTHOR{Jian and Henderson}

\RUNTITLE{Estimating the Probability of Convexity}

\TITLE{Estimating the Probability that a Function Observed with Noise is Convex}

\ARTICLEAUTHORS{%
\AUTHOR{Nanjing Jian, Shane G. Henderson}
\AFF{School of Operations Research and Information Engineering, Cornell University}
\AFF{\EMAIL{nj227@cornell.edu, sgh9@cornell.edu}}
} 

\ABSTRACT{%
Consider a real-valued function that can only be observed with stochastic noise at a finite set of design points within a Euclidean space. We wish to determine whether there exists a convex function that goes through the true function values at the design points. We develop an asymptotically consistent Bayesian sequential sampling procedure that estimates the posterior probability of this being true. In each iteration, the posterior probability is estimated using Monte Carlo simulation. We offer three variance reduction methods -- change of measure, acceptance-rejection, and conditional Monte Carlo. Numerical experiments suggest that the conditional Monte Carlo method should be preferred.
}%



\maketitle

%

\section{Introduction}
\label{sec:intro}
Our goal in this paper is to develop a method to determine whether a real-valued function $g: S \subseteq \mbR^d \rightarrow \mbR$, observed at a finite set of points $\xv_1, \xv_2, \ldots, \xv_r$ in $S$ with noise from a stochastic simulation model, is convex in the sense that a convex function $f$ exists that coincides with $g$ at $\xv_1, \xv_2, \ldots, \xv_r$. That is, does there exist a convex function $f$ that goes through the points $(\xv_1, g(\xv_1)), (\xv_2, g(\xv_2)), \ldots, (\xv_r, g(\xv_r))$? For example, $g$ could be the expected profit in an inventory problem where the demand $\xi$ is random, the starting inventory $\xv$ of the day can take integer values $S = \{0, 1, \ldots, \infty\}$, and we only observe a simulation estimate of $g$ at $\xv$. We might choose a few integer values in $S$ to test whether the expected profit is convex as a function of the starting inventory. Or $g$ might represent the (true) expected waiting time for an ambulance as a function of base locations represented in latitude-longitude coordinates, where we observe $g$ with noise through a stochastic simulation model of ambulance operations. We might then choose a finite set of base-location options in the city to test whether the waiting time is convex with regard to the location of the ambulance bases.

Convexity is a key structural property that can be exploited in many ways. Assuming the minimum is attained, one can use gradient-based methods (for smooth functions) or a cutting-plane based method (for nonsmooth functions) to quickly find the minimum, or bounds on the minimum, e.g., \cite{nes04,Glynn2013}. Even if a function is not globally convex, one might use our methodology to identify regions around local minima in which the restriction of the objective function is convex (``basins of attraction''). Such basins can provide information on the stability of a solution \citep{vog88}.
	
Our methodology is computationally demanding, so we do not envisage it being used only once prior to the selection of an optimization algorithm, which is then applied to solve an optimization problem only once. Rather, we contend that it is usually the case that optimization models are repeatedly solved, often with the same model structure but with different data. In such cases, one would expect that the presence of convexity (or not) would usually be preserved from one data set to another. So then we believe it appropriate to explore convexity {\em once} on {\em one} data set using our methodology, and to use the results to inform subsequent effort on {\em many} instances of the optimization problem that differ in modest ways, perhaps in terms of input parameters or problem-specific data. In this view, the computational cost of exploring the function once with our methodology is amortized over the subsequent analysis.

Beyond optimization, convexity can also provide insights into qualitative model behavior in simulation applications and elsewhere. This is especially useful when we have a sequence of similar simulation models with different inputs but the same structural properties, as mentioned earlier.

Most studies on convexity detection use a frequentist hypothesis-testing framework, and can be categorized with regard to how the null hypothesis is defined. The first category uses an infinite-dimensional functional approach. It defines the null hypothesis as $g \in \mcC$, where $\mcC$ is the cone of convex functions in an appropriate function space, whereas the alternative hypothesis is $g \notin \mcC$. The representative paper \cite{JN2002} models $g$ as a Gaussian process assuming smoothness under the null hypothesis, and uses the $L^r$ distance between $g$ and $\mcC$ as the test statistic.

The second category is closer to our paper in that it works with finite-dimensional vectors. The null hypothesis is $\g \in \mbC$, where $\g$ is the restriction of $g$ to a finite set of points, and $\mbC$ is the set of convex functions restricted to the same set of points. In this case, the noisy evaluations of $\g$ are modeled by a Gaussian random vector. Although not targeted specifically for testing convexity, \cite{Silvapulle2001} describes a general approach for testing whether a finite-dimensional Gaussian vector lies within a closed convex cone. They use the distance between the Gaussian vector and the cone as the test statistic, and show that this distance follows a so-called chi-bar-square distribution, the tail probability of which can be evaluated using simulation. 

The third category fits a regression model with Gaussian noise on the observations of $g$ and tests the hypothesis of convexity through the estimated model parameters. This approach is essentially testing whether there exists a convex function that could have generated the observed finite set of function values. In one dimension and under regularity conditions, \cite{BHL2005} show that testing for the regression parameter is equivalent to testing $\g \in \mbC$, when $\mbC$ is defined with regard to nonnegative second order Vandermonde determinants. Their test is based on the idea that if a one-dimensional function is convex, then the sample mean of the function values in a partition should be lower than certain linear combinations of the function values in neighboring partitions. Others fit cubic splines on the observations and use the second order derivatives at the knots to test for convexity, e.g. \cite{DT1998}, \cite{Wang2011} and \cite{Meyer2012}. For higher dimensions, \cite{Abrevaya2005} work with small and localized sets of data points and count all the possible convex and concave simplices to construct a test statistic. \cite{Lau1978} uses a second-order parametric model for the data points, and is a good survey of early literature. 

A closely related field to convexity tests is convex regression, where one fits a regression model to observations under the constraint that the fitted model is convex. Work in this direction includes \cite{Judge1966}, \cite{Allon2007}, \cite{Seijo2010}, \cite{Hannah2011}, \cite{Lim2012}, among which \cite{Seijo2010} provides a review of past work. 

In this paper, we give a Bayesian sequential algorithm that iteratively collects noisy evaluations of an unknown function $g$ on a fixed and finite set of design points $\xv$, and uses them to estimate the posterior probability that the function, when restricted to the design points, is convex. Our approach differs from previous research in two main ways. First, since function estimates are obtained via Monte Carlo simulation, we can only observe the function $g$ on a finite set of points $\xv$. Thus the best we can provide is a statistical guarantee that there exists a convex function that coincides with the unknown function $g$ at those points. Even if there exists such a convex function, it does not imply that $g$ itself is convex, although nonexistence does imply nonconvexity of $g$. Second, we use a Bayesian conjugate prior model that updates a posterior on the function values every time we collect a set of new samples. Then the posterior probability of convexity is estimated separately through Monte Carlo simulation. Instead of having a fixed running time, as is the case with hypothesis testing, our algorithm can be stopped at any stage to output an estimated probability of convexity. Indeed, the Bayesian approach avoids the difficulty in frequentist sequential hypothesis testing where one should condition on the outcomes of previous hypothesis tests when considering the distribution of a test statistic. The Bayesian framework seems to us to be more straightforward in both concept and implementation.

Our overall approach is to successively update a posterior distribution on the vector of (true) function values $\g$. In doing so we assume that the noise in the estimated function values is normally distributed and adopt a conjugate prior so that we can use standard posterior updates. In simulation the normality assumption is common and usually reasonable, since one can average multiple (finite variance) replications to obtain a single approximately normally distributed replication through the central limit theorem. Normality is not essential to our approach; other distributions could be assumed, but the use of a conjugate prior is key to keeping the computation manageable. We restrict attention to the normal assumption for brevity and simplicity, and because our primary intended application area is in simulation, where the assumption can almost always be at least approximately satisfied (through averaging).

For a given posterior distribution, computing the probability of convexity for a sample from the posterior appears to be difficult. We use Monte Carlo to estimate this probability, providing three methods for reducing the variance of the Monte Carlo estimator of the posterior probability of convexity. The change of measure and acceptance-rejection methods reuse samples obtained in an earlier iteration to construct an unbiased estimator for the current iteration. These two methods can be useful in any sequential algorithm in a Bayesian framework, but need to be used with caution due to heavy-tailed behavior of the likelihood ratio that is needed in these methods. The conditional Monte Carlo method takes advantage of the spherical property of Gaussian and $t$ distributions. It can be applied to the more general problem of estimating the probability that a Gaussian or $t$-distributed random vector lies in a polyhedron, which might arise, e.g., in solving linear feasibility problems described in \cite{Szechtman2016}.

We view the Bayesian methodology developed herein, where we repeatedly estimate the posterior probability of convexity as $n$, the number of simulation replications, increases, as being well suited to exploratory analysis aimed at developing knowledge of the structure of the function $g$.  Certainly, a key strength of the Bayesian approach, relative to hypothesis tests, is that it allows an analyst to explore several values of $n$, interactively increasing $n$ until satisfied with the results. Hypothesis tests do not afford this flexibility, so in our opinion are not as well suited to the exporatory analysis we envisage. Our statistical guarantees are weaker than those of formal hypothesis tests, in that the confidence intervals we generate on the posterior probability of convexity hold only marginally in $n$ and not {\em jointly} in $n$ (they are not confidence bands). However, plotting the confidence intervals as a function of $n$, as we do, provides a visual sense of whether the function $g$ (restricted to the selected points) is convex or not, along with a sense of the sample size needed to establish convexity or not with some reliability. It is conceivable that one could further develop the Bayesian methodology to develop a more rigorous test, but we do not pursue that line, again because of our focus on an exploratory tool.

This paper is built upon \cite{Jian2014}, with the addition of new results, complete proofs, new variance reduction methods, and more extensive numerical results. Full proofs of several results can be found in the online supplement.

\paragraph{Notation:} We use upper case Latin letters for random variables or sets, and lower case Latin letters for deterministic variables. Vectors are in bold, and matrices are in upper case Greek letters. We use $A^T$ to denote the transpose of the matrix $A$. For a set $S$, $S^{\circ}$ is its interior. We use $\inD$ for convergence in distribution, and $\ind{B}$ is the indicator function that takes the value $1$ on the event $B$ and $0$ otherwise.

\section{Problem Statement and Assumptions}
\label{sec:notation}

Suppose we can obtain noisy evaluations of a real-valued function $g: S \rightarrow \mbR$ over a fixed and finite set of design points $\xv=\{\xv_i: i = 1, 2, \ldots, r \}$ in $S \subseteq \mbR^d$. Ideally, we would like to know whether the function $g$ is convex on $S$ or not. In the absence of any regularity assumptions on $g$, such as are assumed in \cite{JN2002}, it appears that this question cannot be addressed in finite time. 
The rest of this paper is focused on giving a statistical guarantee on the convexity of the finite-dimensional vector obtained by restricting the function $g$ to the $r$ points in $\xv$, i.e., we restrict attention to $\g = (g(\xv_1),g(\xv_2),\ldots,g(\xv_r)) \in \mbR^r$. 

\begin{definition}\label{def:vectorConvex}
Given a finite set of points $\xv$, we define a vector $\g$ to be convex if and only if there exists a convex function $g$ whose values on $\xv$ coincide with $\g$, i.e. $g(\xv) = \g$. 
\end{definition}
\begin{definition}\label{def:coneC}
Define $\mbC = \mbC(\xv_1,\xv_2,\ldots,\xv_r) \subseteq \mbR^r$ to be the set of all convex vectors on $\xv = (\xv_1,\xv_2,\ldots,\xv_r)$, so that $\g$ is convex if and only if $\g \in \mbC$.
\end{definition}
The notion of vector convexity is weaker than (the usual) functional convexity, since if a function $g$ is convex then its restriction $\g$ on $\xv$ is convex under our definition. The converse is not true since we can arbitrarily extend $\g$. 

The set $\mbC$ is a convex cone. In Section~\ref{sec:convexity}, we will see that $\g \in \mbC$ if and only if a certain linear system is feasible, and the linear system is then a tool to verify vector convexity. We will also show that $\g$ is strictly convex (in the sense that a strictly convex function $g$ exists that agrees with $\g$ on $\xv$) iff $\g \in \intC$, the interior of $\mbC$, in the online supplement.

We use a Bayesian approach, regarding $\g$ as an unknown realization from the prior distribution of a random vector $\f$. Let $(\bm{\xi}_j: j =1, 2, \ldots)$ be an i.i.d.\ sequence of $r$-dimensional Gaussian random vectors, each with mean vector $\bm{0}$ and covariance matrix $\Gamma$, that is independent of $\f$. Our $j$th observation is then $\Y_j$, where $\Y_j = \f + \bm{\xi}_j$, $j = 1, 2, \ldots$. We denote the $i$th component of the vector $\Y_j$ by $Y_{ij}$, and interpret it as the $j$th sampled function value at the point $\xv_i$.

To elaborate, we view $\g$ as a deterministic vector of function
  values. In the Bayesian structure within which we work, the vector $\g$ is
 modeled through a random vector $\f$ that is sampled from a prior distribution at the outset, and we subsequently accumulate evidence on the value of $\f$ (and thus $\g$) through the samples $\Y_1, \Y_2, \ldots$, updating our beliefs through the posterior distribution. In principle, with sufficient replications we can recover the value of $\f$; our progress towards this goal is reflected through the posterior distribution. 
 
We observe the value of $\f$ with additive noise, through the vector outputs of successive simulation replications $\Y_1, \Y_2, \ldots$. Consistent with much simulation literature, conditional on $\f$, the observed values $\Y_j, j = 1, 2, \ldots$ are assumed to be Gaussian, which can at least be approximately satisfied by averaging multiple simulation replications to yield a single $\Y_j$. This is an approximation, but it affords considerable computational advantages, as we will see later.

The covariance matrix $\Gamma$ is not necessarily diagonal, i.e., we do not necessarily constrain the observations to be (conditionally) independent between sampled points, conditional on $\f$. Thus, Common Random Numbers (CRN) can be employed within our framework. CRN induces positive correlation on the $r$ dimensions of the noise $\bm{\xi}$, so that the structure of the underlying ``true'' function $\f$ can be better preserved than would be possible with conditionally independent observations \citep{cheanknel12}. For simplicity, we assume throughout that the covariance matrix $\Gamma$ is positive definite.

\section{Sequential Algorithm}\label{sec:algorithm} 
Initially, before any sampling, we fix the $r$ points $\xv_i, i = 1, \ldots, r$ and the prior mean $\bm{\mu_0}$ and covariance matrix $\Lambda_0$ of the assumed Gaussian prior distribution of $\f$. At the beginning of the $n$th iteration $(n = 1, 2, \ldots)$, we obtain a new observation $\Y_n$, and use that to update the posterior distribution on the function values, as described in Section~\ref{sec:updates}. The information collected thus far is denoted $\mathscr{A}_n$, which is the sigma field generated from $\mathscr{A}_0$ and $\{\Y_j, j=1,\ldots,n\}$. Thus, $\{\mathscr{A}_n\}_{n=0, 1, 2, \ldots}$ is a filtration. Once the posterior distribution has been updated, we separately estimate the posterior probability of convexity, $P(\f \in \mbC | \mathscr{A}_n)$ as discussed in Section~\ref{sec:convexity}. At the end of each iteration, we can choose either to stop, or to continue with the current posterior as the prior of the next iteration. More precisely, the algorithm is as follows.

\begin{algorithm}[H] 
\caption{A sequential method for testing for convexity of the function \label{alg:main}}
\begin{algorithmic}[1]
\REQUIRE The Gaussian prior $P(\f \in \cdot| \mathscr{A}_0)$, with hyperparameters $\{\bm{\mu_0}, \Lambda_0\}$ of the function values $\f$.
\STATE Initialize $n=0$.
\REPEAT 
\STATE Set $n = n+1$.
\STATE Obtain a new vector $\y_n$ of $r$ noisy function values at $\xv_1, \xv_2, \hdots, \xv_r$.
\STATE Update the posterior distribution of $\f|\mcA_n$ from the new samples $\y_n$ using $\f|\mcA_{n-1}$ as the prior, as in Section~\ref{sec:updates}.
\STATE Estimate $p_n = P(\f \in \mbC | \mathscr{A}_n)$ from the distribution of $\f | \mathscr{A}_n$ using the Monte Carlo method described in Section~\ref{sec:convexity}, obtaining a confidence interval ${[ \hat{p}_n - h_n, \hat{p}_n + h_n ]}$.
\UNTIL{stopped}
\RETURN A confidence interval $[ \hat{p}_n - h_n, \hat{p}_n + h_n ]$ of $p$.
\end{algorithmic}
\end{algorithm}

In Step 6, the posterior probability that $\f$ is convex is estimated using Monte Carlo simulation by sampling $m$ times from the posterior distribution $\f|\mcA_n$. Compared to obtaining the samples $\Y_j$, which involves running the full simulation model, sampling from the posterior distribution is computationally inexpensive, entailing sampling from a Gaussian or $t$ distribution, depending on whether the variance is known or unknown. Depending on the computational cost of running the full simulation model, we can also skip the estimation of $p_n$ for some $n$ and enter Step 6 only for selected values of $n$.
 
\section{Posterior Updates}\label{sec:updates}

We use a conjugate prior to update our belief about $\f|\mcA_n$ in each iteration $n$. Since we assumed that $\Y - \f \sim N(\bm{0}, \Gamma)$, this conjugate prior is normal-normal when $\Gamma$ is known, and normal-inverse-Wishart when $\Gamma$ is unknown. In this section we give the updating formula of the posterior distribution of $\f|\mcA_n$ in Step~5 of Algorithm~\ref{alg:main} under these two scenarios, given the prior $\f|\mcA_{n-1}$. The formulae given here are standard; see, e.g., \cite{DeGroot1970}, \cite{Gelman2003}, or \cite{Bernardo2008}. 

\subsection{Conjugate Prior under Known Sampling Variance}
\label{sec:update_knownvar}

First, before any sampling we select a non-informative Gaussian prior with zero mean and large variance, i.e. $\f | \mcA_0 \sim N(\bm{\mu_0},\Lambda_0)$ in which $\mu_0 = \bm{0} \in \mbR^r$ and $\Lambda_0 \in \mbR^{r \times r}$ is a diagonal matrix with diagonal values that are large relative to the sampling variance (the diagonal of $\Gamma$). Alternative parameters for the Gaussian prior could be used in the presence of more information, and would not change the algorithm.

At iteration $n > 1$, the posterior from the last iteration $\f|\mathscr{A}_{n-1} \sim N(\bm{\mu_{n-1}},\Lambda_{n-1})$ is used as the prior for the current iteration. We then obtain $s \ge 1$ new objective-function samples $(y_{ij}, j= 1, 2, \ldots, s)$ at each of the design points $\xv_i, i = 1, 2, \ldots, r$. The mean $\bm{\mu_n}$ and covariance $\Lambda_{n}$ of the posterior $\f|\mathscr{A}_{n} \sim N((\bm{\mu_{n},\Lambda_{n}})$ are updated by
\begin{align}
\Lambda_n^{-1} &= \Lambda_{n-1}^{-1} + s \Gamma^{-1} \notag \\
\bm{\mu_n} &=\Lambda_n (\Lambda_{n-1}^{-1} \bm{\mu_{n-1}} + s \Gamma^{-1}\bar{\y}), \label{eqn:knownVarPosterior}
\end{align}
where the $i$-th component of the $r$-dimensional vector $\bar{\y}$ is $s^{-1}\sum_{j=1}^{s} \y_{ij}$. One can adaptively choose the sample size $s$ in each iteration, but for simplicity we use $s=1$, meaning that only one new sample is obtained in each iteration.

The updating equation \refeq{eqn:knownVarPosterior} involves matrix inversion. To reduce the computational effort, we use Cholesky factorization and the Sherman-Morrison-Woodbury formula as detailed in the online supplement.

\subsection{Conjugate Prior under Unknown Sampling Variance}
\label{sec:update_unknownvar}

When the sampling variance $\Gamma$ is unknown, the inverse-Wishart distribution provides a conjugate prior. First, we use an uninformative Jeffrey's prior, where the prior joint distribution on $\f$ and $\Gamma$ is proportional to $|\Gamma|^{-(r+1)/2}$ \citep{Gelman2003} and $|A|$ denotes the determinant of the matrix $A$. To construct Jeffrey's prior, an initial set of $r$-dimensional samples $\y_j = (y_{ij}, i = 1, 2, \ldots, s), j = 1,2,\ldots,s_0$ are used to estimate the parameters of the normal distribution for the mean $\f$ and the Inverse-Wishart distribution (Inv-Wishart) for the variance $\Gamma$. The initial sample size $s_0$ can be any positive integer. For a prior that reflects the data without being too costly, we choose $s_0 = r+1$, where $r$ is the number of design points. This choice also ensures that the inverse-Wishart distribution is concentrated on covariance matrices that are positive definite. More specifically \citep{Gelman2003}, 
\begin{equation}
\begin{aligned}
\Gamma | \mcA_0, \y &\sim \text{Inv-Wishart}_{\upsilon_0}(\Xi_0^{-1}) \\
\f | \Gamma, \mcA_0, \y &\sim N(\mu_0, \Gamma/\kappa_0), \\
\end{aligned} \label{eqn:unknownVarPrior}
\end{equation}
where 
\[
\bm{\mu_0} = \frac{1}{s_0}\sum_{j=1}^{s_0} \y_j = \bar{\y}; \quad
\kappa_0 = s_0; \quad
\upsilon_0 = s_0-1; \quad
\Xi_0 = \left(\sum_{j=1}^{s_0} (\y_j - \bar{\y})(\y_j - \bar{\y})^T\right)^{-1}. 
\]

In iteration $n \ge 1$, we obtain $s$ samples $y_{ij}$ on each of the points $\xv_i, i = 1, \ldots, r$ and update the posterior of 
\begin{equation}
\begin{aligned}
\Gamma | \mcA_n &\sim \text{Inv-Wishart}_{\upsilon_n}(\Xi_n^{-1}) \\
\f | \Gamma, \mcA_n &\sim N(\bm{\mu_n}, \Gamma/\kappa_n)
\end{aligned}
\end{equation}
by \citep{Gelman2003}:
\begin{equation}
\begin{aligned}\label{eqn:unknownVarPosterior}
\bm{\mu_n} & = \frac{\kappa_{n-1}}{\kappa_{n-1} + s}\bm{\mu_{n-1}} + \frac{s}{\kappa_{n-1} + s} \bar{\y}; \quad
\kappa_n = \kappa_{n-1} + s; \quad
\upsilon_n = \upsilon_{n-1} + s; \\
\Xi_n & = \Xi_{n-1} + S + \frac{\kappa_{n-1} s}{\kappa_{n-1} + s} (\bar{\y} - \bm{\mu_{n-1}})(\bar{\y} - \bm{\mu_{n-1}})^T, 
\end{aligned}
\end{equation}
where the $i$-th component of the $r$ dimensional vector $\bar{\y}$ is defined as $\bar{\y}_i = \sum_{j=1}^{s} y_{ij}/s, \; i = 1, \ldots, r$, and the $r \times r$ matrix $S$ is the sum of squared errors $\sum_{j=1}^{s} (\y_j - \bar{\y})(\y_j - \bar{\y})^T$. For simplicity we again choose $s=1$ when updating, so that $\bm{y}_j = \bar{\bm{y}}$ and $S=0$.

If a random $r \times r$ matrix $\Gamma$ has the Inverse-Wishart distribution with parameters $\upsilon$ and $\Xi^{-1}$, whose density is proportional to $|\Xi|^{\upsilon/2} |\Gamma|^{-(\upsilon-r-1)/2} \exp\{-\mbox{tr}(\Xi \Gamma^{-1})/2\}$, where $\mbox{tr}(\cdot)$ is the trace of a matrix, the inverse $\Gamma^{-1}$ has the Wishart distribution with parameters $\upsilon$ and $\Xi$. The Wishart distribution is a higher-dimensional generalization of the $\chi^2$ distribution, and thus can be expressed similarly as the sum of squares of Gaussian random vectors. To generate a $\text{Wishart}_{\upsilon}(\Xi)$ distributed random matrix $\Gamma$, we generate $\upsilon > r$ independent, $r$-dimensional random vectors $W_i$ distributed as $N(0, \Xi)$, and return $\Gamma = \sum_{i=1}^{\upsilon} W_i W_i^T$.

With the posterior covariance distributed as Inverse-Wishart and the posterior mean distributed as Gaussian conditioning on the covariance, the marginal distribution of the posterior mean $\f | \mcA_n$ is
\begin{equation}\label{eqn:unknownVarPosteriorMarginal}
\f | \mcA_n \sim t_{(\upsilon_n - r + 1)}(\bm{\mu_n}, \Xi_n/ (\kappa_n(\upsilon_n - r + 1))),
\end{equation}
where $t_{(\upsilon_n - r + 1)}(\bm{\mu_n}, \Xi_n/(\kappa_n(\upsilon_n - r + 1)))$ is a multivariate Student-$t$ distribution with $(\upsilon_n - r + 1)$ degrees of freedom, location parameter $\bm{\mu_n}$, and scale matrix $\Xi_n/(\kappa_n(\upsilon_n - r + 1))$. The density function of $\f|\mcA_n$ is thus proportional to $|\Xi_n/(\kappa_n(\upsilon_n - r + 1))|^{-1/2} \{1+ (\f-\bm{\mu_n})^T [\Xi_n/(\kappa_n(\upsilon_n - r + 1))]^{-1}(\f-\bm{\mu_n})\}^{-(\upsilon_n+1)/2}$ \citep{Gelman2003}. One can generate such a random vector by exploiting the elliptical nature of the distribution; see Section~\ref{sec:conditionalMC}. (Although that section does not explicitly give a generation algorithm, an algorithm should be clear from the arguments given there.)

\section{Convexity}
\label{sec:convexity}
Recall that $\g = (g(\xv_1), g(\xv_2), \ldots, g(\xv_r))$, and the vector $\g$ is defined to be convex if and only if there exists a convex function that coincides with $\g$ on the set of points $\xv_1, \ldots, \xv_r$. Equivalently, for each $i = 1, \ldots, r$ there exists a hyperplane $\{\bm{a}_i^T \xv + b_i: \xv \in \real^d\}$ that goes through $(\xv_i, g(\xv_i))$ and lies at or below all the other points $(\xv_j, g(\xv_j)), j \neq i$ (\citealt[p.539]{mur88}; \citealt{atlepehen03}). That is, $\g$ is convex if and only if there exists feasible solutions $\bm{a}_i \in \mbR^{d}, i = 1, \ldots, r$ and $\bm{b} \in \mbR^{r}$ to the linear system
\begin{equation} \label{LS:big}
\begin{aligned}
& \bm{a}_i^T \xv_i + b_i = g(\xv_i), \; \mbox{ for all } i \in \{1,\ldots,r\}\\
& \bm{a}_i^T \xv_j + b_i \leq g(\xv_j), \;\mbox{ for all } i \in \{1,\ldots,r\} \mbox{ and } j\neq i \mbox{, }  j \in \{1,\ldots,r\},
\end{aligned} \tag{LS}
\end{equation} with $b_i$ being the $i$-th component of $\bm{b}$. Let the set of all $\g$ such that the corresponding \ref{LS:big} is feasible be $\mbC$, which denotes the set of all convex vectors $\g$ with regard to the $r$ design points $\xv_1, \ldots, \xv_r$.

This large linear system can also be decomposed into $r$ sub-systems, indexed by $i = 1, \ldots, r$:
\begin{equation} \label{LS:infeasDecomp}
\begin{aligned}
& \bm{a}_i^T \xv_i + b_i = g(\xv_i) \;\\
& \bm{a}_i^T \xv_j + b_i \leq g(\xv_j), \;\mbox{ for all } j\neq i \mbox{ and }  j \in \{1,\ldots,r\},
\end{aligned} \tag{LS$(i)$}
\end{equation}
each with the variables $\bm{a}_i \in \mathbb{R}^d$ and $b_i \in \mathbb{R}$.

Transforming the question of whether a vector is convex to the feasibility of $r$ linear systems allows us to use Monte Carlo simulation to estimate the posterior probability of convexity at the end of each iteration $n$. We first simulate $m$ random samples from the posterior distribution of $\f | \mcA_n$. Then, for each generated sample, we determine feasibility (or lack thereof) for the linear systems (\ref{LS:infeasDecomp}, $i = 1, \ldots, r$) in sequence. If {\em any} linear system is infeasible then we stop (skip the rest of the systems) and conclude that this generated sample is not convex, since one cannot define an appropriate hyperplane. The probability $P(\f \in \mbC | \mathscr{A}_n)$ is then estimated by the sample average of the indicators of convexity for each sample as described more formally in Algorithm~\ref{alg:cvx}.

\begin{algorithm}[htb] 
\caption{Subroutine used in Step 6 of Algorithm \ref{alg:main} to estimate $P(\f \in \mbC | \mcA_n)$.\label{alg:cvx}}
\begin{algorithmic}[1]\label{alg:pconvex1}
\REQUIRE The posterior marginal density of $\f | \mcA_n$ from \refeq{eqn:knownVarPosterior} or \refeq{eqn:unknownVarPosteriorMarginal}.
\STATE Generate independent samples $\{\y_n^1, \y_n^2, \ldots, \y_n^m\}$ from the posterior marginal density of $\f | \mcA_n$.
\FOR{$k$ from 1 to $m$}
\STATE Set $\ind{\y_n^k \in \mbC} = 1$.
\FOR{$i$ from 1 to $r$}
\STATE Set $g(\xv_i)$ as the $i$-th component of $\y_n^k$, $i = 1, \ldots, r$.
\STATE Solve for the feasibility of \ref{LS:infeasDecomp}. 
\IF{\ref{LS:infeasDecomp} is infeasible} 
\STATE Set $\ind{\y_n^k \in \mbC} = 0$.
\STATE BREAK the inner loop and go to next $k$.
\ENDIF
\ENDFOR
\ENDFOR
\RETURN The center $\hat{p}_n = \sum_{k=1}^m \ind{\y_n^k \in \mbC} / m$ and half-width $h_n = 1.96s_n/\sqrt{m}$ of a 95\% confidence interval for $P(\f \in \mbC | \mcA_n)$, where $s_n$ is the sample standard deviation of $\ind{\y_n^k \in \mbC}, k = 1, \ldots, m$.
\end{algorithmic}
\end{algorithm}

\section{Asymptotic Validity of the Main Algorithm}\label{sec:asymptotic}

We now establish that the posterior probability of convexity converges to 1 or 0, depending on whether $\g$ is convex or not, with one qualification. If $\g$ is convex but not strictly convex then it lies on the boundary of $\mbC$, and then certain arbitrarily small perturbations of the function values $\g$ will yield points outside $\mbC$. Since we estimate the function values $\g$ using simulation, we cannot rule out such perturbations, and so we should not expect the posterior probability of convexity to converge to 1 or 0.

The formal statement of convergence is with respect to the probability space containing both the prior from which $\f$ is sampled, and the data. We show that when $\f$ is strictly convex the posterior probability of convexity converges to 1, and when $\f$ is not convex the posterior probability of convexity converges to 0. The remaining case where $\f$ lies on the boundary of $\mbC$ has probability 0 under our prior, which has a density with respect to Lebesgue measure.

\begin{theorem}\label{thm:converges}
Let $p_n = P(\f \in \mbC | \mathscr{A}_n)$ be the $n$-iteration posterior probability that $\f$ is convex as in Algorithm~\ref{alg:main}. As the number of iterations $n \rightarrow \infty$, $p_n - \mathds{1}\{\f \in \mbC\} \to 0$ a.s.
\end{theorem}

\cite{Jian2014} has a sketch of the proof in the known variance case. We provide a complete proof that covers both the known and unknown variance cases in the online supplement.

The result of Theorem \ref{thm:converges} relates to the exact posterior probability of convexity, which we estimate using Monte Carlo. We next show that the Monte Carlo estimator from Section \ref{sec:convexity} of the exact probability converges to the same indicator provided that the Monte Carlo sample sizes increase without bound, through a uniform law of large numbers.

\begin{corollary}\label{thm:MCconverges}
Let $p_n^m$ be the $m$-sample estimator of $P(\f \in \mbC | \mathscr{A}_n)$ from Algorithm \ref{alg:pconvex1}. As $n \to \infty$ and $m = m(n) \to \infty$, $p_n^m - \ind{\f \in \mbC} \to 0$ in probability.
\end{corollary}

\proof{Proof.}
We have $|p_n^m - \ind{\f \in \mbC}| \le |p_n^m - p_n| + |p_n - \ind{\f \in \mbC}|$, where $p_n = P(\f \in \mbC | \mathscr{A}_n)$ and $p_n^m = \frac{1}{m} \sum_{k=1}^{m} \mathds{1}\{\y_n^k \in \mbC \}$. Let $\epsilon > 0$ be arbitrary. For the first term, Chebyshev's inequality gives
\begin{align*}
P(|p_n^m - p_n| > \epsilon) &=  E P\left(\left|\frac{1}{m} \sum_{k=1}^{m} \mathds{1}\{\y_n^k \in \mbC \} - P(\f \in \mbC | \mathscr{A}_n)\right| > \epsilon \Big| \mathscr{A}_n\right)\\
& \le  E \left( \frac{Var(\mathds{1}\{\y_n^k \in \mbC \}|\mcA_n)}{m \epsilon^2} \right)
 \le \frac{1}{4m\epsilon^2} \to 0
\end{align*}
as $n \rightarrow \infty$ since $m = m(n) \to \infty$ as $n \to \infty$. This shows that $p_n^m - p_n \to 0$ in probability as $n \to \infty$. Also Theorem \ref{thm:converges} shows that $|p_n - \ind{\f \in \mbC}| \to 0$ in probability as $n \to \infty$.

\section{Variance Reduction Methods}
\label{sec:varianceReduction}
In this section, we improve the vanilla Monte Carlo method through three variance-reduction methods. The change of measure and acceptance-rejection methods are likelihood-ratio-based methods that reuse samples generated in an earlier iteration, and the conditional Monte Carlo method reduces the variance through smoothing.

\subsection{Change of Measure}
\label{sec:changeofmeasure}
Algorithm \ref{alg:pconvex1} can be computationally costly due to the need to solve up to $mr$ linear feasibility problems \ref{LS:infeasDecomp}, where $m$ is the number of Monte Carlo samples and $r$ is the number of design points. We can reduce the computational effort by reusing samples generated in a previous iteration through a change-of-measure method. The resulting estimator is based on the same principle used in the score-function method for simulation optimization \citep{rubsha90}, and that used in ``green simulation" \citep{Feng2015}. We will see that the resulting estimator is unbiased and has finite variance, but does not perform as well as we might hope. 

Recall that in iteration $n$, Algorithm \ref{alg:pconvex1} generates $m$ i.i.d. samples $\{\y_n^k: k=1,2,\ldots,m \}$ from the posterior marginal distribution of $\f | \mcA_n$ and produces $m$ indicators $\{\ind{\y_n^k \in \mbC}: k=1,2,\ldots,m \}$ of convexity. To reuse these samples, in iteration $n + \ell$, we instead output 
\begin{equation}\label{eqn:changeMeasure}
\hat{p}_{n+\ell} = \frac{1}{m} \sum_{k=1}^m \ind{ \y_n^k \in \mbC } L_{n+\ell,n}(\y_n^k)
\end{equation}
as an estimate of $p_{n+\ell} = P(\f \in \mbC | \mcA_{n+\ell})$, where $L_{n+\ell,n}(\cdot) = \phi_{n+\ell}(\cdot)/\phi_n(\cdot)$ is the likelihood ratio of the densities of $\f|\mcA_{n+\ell}$ and $\f|\mcA_{n}$.

\begin{theorem}\label{thm:finiteVariance}
The change of measure estimator $\hat{p}_{n+\ell} = \mathds{1}\{ \Y_n \in \mbC \} L_{n+\ell,n}(\Y_n)$ is (conditionally) unbiased and has finite conditional variance, conditional on $\mcA_{n + \ell}$ for any $n$ and $\ell \ge 1$.
\end{theorem}

The proof for the known $\Gamma$ case can be found in \cite{Jian2014}, and we provide a proof for the unknown $\Gamma$ case in the online supplement.

Given that the change of measure estimator is unbiased and has finite variance, it is tempting to generate a single sample and re-use it for many iterations to save computational effort. Unfortunately, such an estimator has poor empirical performance. Figure \ref{fig:convex_30d_cm} gives an example where the estimated probability of convexity is greater than 1. This happens especially later in the run when all of the linear systems are feasible, and the likelihood ratios $L_{n+\ell,n}(\y_k^n)$ occasionally take very large values.

Occasional large values of the likelihood ratio $L_{n+\ell,n}(\y)$ might arise when the sample $\y$ is generated within the tail of $\phi_n$. Indeed, Proposition \ref{thm:LRHeavyTail} below shows that in at least one special case, $L_{n+\ell,n}$ has a heavy tail given any sampling trajectory $\mcA_n$. At first sight, this may appear to contradict Theorem~\ref{thm:finiteVariance}, which states that given the posterior information $\mcA_{n+\ell}$ in iteration $n+\ell$, the change of measure estimator is bounded. But notice that in Theorem~\ref{thm:finiteVariance} we are conditioning on more information than in Proposition \ref{thm:LRHeavyTail}. In effect, Proposition~\ref{thm:LRHeavyTail} shows that given the posterior information $\mcA_n$ in iteration $n$, the change of measure estimator in iteration $n+\ell$ could have poor behavior, depending on the (random) samples that are used to update $\f|\mcA_{n+\ell}$ from $\f|\mcA_n$. Thus there is no contradiction between these two results. Proposition~\ref{thm:LRHeavyTail} shows that the change of measure estimator could exhibit volatile behavior when extreme values of the likelihood ratio arise at values that were sampled in iteration $n$. However, we make no claim about how likely such values are to arise. Numerical experiments given later show that indeed the change of measure estimator is volatile.

\begin{proposition}\label{thm:LRHeavyTail}
When $\Gamma$ is known and $r=1$, given $\mcA_n$, $C_n = \sup_{\y \in \mbR^r} L_{n+1,n}(\y)$ asymptotically (as $n \rightarrow \infty$) has the same distribution as $e^{\chi^2_1}$, where $\chi^2_1$ is a non-central chi-square random variable with 1 degree of freedom.
\end{proposition}

In fact, the proof for Proposition~\ref{thm:LRHeavyTail} in the online supplement also applies when $n$ is finite. In that case $\ln C_n$, conditional on $\mcA_n$, is a non-central chi-square random variable scaled by a constant of order $O(1/n)$ and shifted by another constant of order $O(1/n)$. The conclusion of Proposition~\ref{thm:LRHeavyTail} can be generalized to $r>1$ when $\Gamma$ is known and diagonal. Indeed, when $\Gamma$ is diagonal the likelihood ratio decomposes into a product, so that $\ln C_n = \ln (\sup_{\y \in \mbR^r} L_{n+1,n}(\y)) = \sup_{\y \in \mbR^r} \ln (\prod_{i=1}^{r} L_{n+1,n}(\y_i)) = \sum_{i=1}^{r} \sup_{\y_i \in \mbR} \ln L_{n+1,n}(\y_i)$. Proposition~\ref{thm:LRHeavyTail} then allows us to conclude that, conditional on $\mcA_n$, this is asymptotically conditionally distributed as $\chi_r^2/2$, where $\chi_r^2$ is a non-central chi-square random variable with $r$ degress of freedom.
Since the tail probability of $\chi_r^2/2$ at a given point increases in $r$, we expect this heavy tail behavior to be more significant as $r$ increases, i.e., as the number of design points increases. We conjecture that the likelihood ratio is similarly heavy-tailed in the cases where $\Gamma$ is known but not necessarily diagonal and when $\Gamma$ is unknown.


In summary, conditional on $\mcA_{n+\ell}$, the estimator $\hat{p}_{n+\ell}$ is unbiased and has finite variance, but its distribution may be heavy tailed given $\mcA_n$ only, depending on the samples obtained to update to $\f|\mcA_{n+\ell}$. Thus this estimator needs to be used with caution. We suggest that if the method is to be used, then one should do so with small $\ell$, e.g., $\ell < 5$, based on simulation experiments described later.

\subsection{Acceptance/Rejection} 
\label{sec:AR}
The change of measure estimator reuses all the samples obtained in an earlier iteration by outputting a Monte Carlo estimator that scales each indicator $\{I_n^k = \mathds{1}\{\Y_n^k \in \mbC\}: k=1,2,\ldots,m \}$ by a likelihood ratio $L_{n+\ell,n}(\Y_n^k) = \phi_{n+1}(\Y_n^k)/\phi_n(\Y_n^k)$, where $\phi_n$ is the posterior density. An alternative is to reuse a subset of the samples from the previous iteration through acceptance-rejection.

Suppose that in iteration $n$, we have $m$ i.i.d. Monte Carlo samples $\{\y_n^k: k=1,2,\ldots,m \}$ from $\f|\mcA_n$, together with the indicators $\{I_n^k = \mathds{1}\{\y_n^k \in \mbC\}: k = 1, 2, \ldots, m\}$. Then, at iteration $n+1$, the $k$-th sample $\y_k^n$ will be accepted (reused) with probability $L_{n+1,n}(\y_k^n)/c$, where $c \ge \sup\{L_{n+1,n}(\y): \y \in \mbR^r\}$. If the accepted indices are $A \subseteq \{1,2,\ldots,m\}$, then $m-|A|$ additional samples can be generated from $\f|\mcA_{n+1}$ to ensure a total of $m$ samples. The estimator is then just the usual Monte Carlo estimator based on all $m$ samples, i.e., $\hat{p}_{n+1} = \left( \sum_{k \in A} \ind{\y_k^n \in \mbC} + \sum_{k=1}^{m-|A|} \ind{\y_k^{n+1} \in \mbC}\right)/m$.

When the sampling variance $\Gamma$ is known, optimization shows that $c$ is given by
$$\left\{ \frac{|\Lambda_{n}|}{|\Lambda_{n+1}|} \exp \left[ (\Lambda_{n+1}^{-1}\bm{\mu_{n+1}} - \Lambda_{n}^{-1}\bm{\mu_{n}})^{T} \Gamma (\Lambda_{n+1}^{-1}\bm{\mu_{n+1}} - \Lambda_{n}^{-1}\bm{\mu_{n}}) + \bm{\mu_n}^{T}\Lambda_n^{-1}\bm{\mu_n} - \bm{\mu_{n+1}}^{T}\Lambda_{n+1}^{-1}\bm{\mu_{n+1}} \right] \right\}^{1/2},$$
where the parameters $\bm{\mu_n}, \Lambda_n, \bm{\mu_{n+1}}, \Lambda_{n+1}$ are defined as in Section \ref{sec:algorithm}.
When $\Gamma$ is unknown, $c$ is the maximum of a ratio of polynomials and does not have a closed form, so we calculate it numerically.

The acceptance-rejection estimator is simply an average of i.i.d.\ samples, like the pure Monte Carlo estimator. The difference lies in how the samples are obtained. The probability of accepting a sample generated in iteration $n$ is $1/c$, so the efficiency of this method is related to the constant $c$. According to Proposition \ref{thm:LRHeavyTail}, the likelihood $L_{n+1,n}(y)$ can take very large values, meaning that $c$ can often be large. When $c$ is large, very few of the earlier samples might be reused, so the majority of the $m$ samples needed in the $(n+1)$th iteration are new. This may lower the efficiency of the acceptance-rejection method.

\subsection{Conditional Monte Carlo}
\label{sec:conditionalMC}
Denote the upper hemisphere of the $(r-1)$ spherical shell, $\{z \in \real^r: \|z\|=1, z_r \ge 0\}$ by $S_+^{r-1}$. We can view a sample from the posterior distribution as consisting of both a direction $Z$ chosen from $S_+^{r-1}$ and a step size $T$ taking both positive and negative values along that direction, along with the linear transformation to obtain the appropriate scale matrix and then a translation by the mean. We condition on the direction $Z$, and integrate the posterior over the interval of step sizes $[t_{\text{min}}, t_{\text{max}}]$ that yield points inside the convexity cone $\mbC$. Averaging the results over a number of uniformly generated directions gives the desired estimator.

For convenience, let $E_n(\cdot) = E(\cdot | \mcA_n)$ and $P_n(\cdot) = P(\cdot | \mcA_n)$. We write $X = TZ$, so that in the known variance case, $X \sim N(0,I)$, and in the unknown variance case $X \sim t_{\nu_n}(0,I)$. Then
\begin{align*}
P(\f \in \mbC | \mcA_n)
& = E_n \left(\ind{\Lambda_n^{1/2}X + \bm{\mu_n} \in \mbC} \right)\\
& = E_n \left( \ind{T\Lambda_n^{1/2}Z + \bm{\mu_n} \in \mbC} \right), \text{ for } Z \text{ uniform on } S_+^{r-1} \\
& = E_n \left( E_n \left(\ind{T\Lambda_n^{1/2}Z + \bm{\mu_n} \in \mbC} | Z \right) \right) \\
& = E_n (P_n(T \in [\ti(Z), \ta(Z) ] \; | Z)) \\
& = E_n (F_{T|Z}(\ta(Z)) - F_{T|Z}(\ti(Z))).
\end{align*}
Here $F_{T|Z}$ is the conditional distribution function of $T$ given $\mcA_n$ and $Z$. (We shall see that $T$ is independent of $Z$.) Thus, the posterior probability $P(\f \in \mbC | \mcA_n)$ can be estimated using $F_{T|Z}$ and a way to calculate $\ta(Z)$ and $\ti(Z)$. Theorem \ref{thm:T|Z} gives the former, and linear programs \ref{LS:infeasDecomp} (below) give the latter. 

\begin{theorem}[Distribution of $T|Z$]\label{thm:T|Z}
When the sampling variance $\Gamma$ is known, $F_{T|Z}(t) = (1+\text{sign}(t)F_{\chi^2_r}(t^2))/2$, where $F_{\chi^2_r}(\cdot)$ is the (cumulative) distribution function of a $\chi^2$ r.v.\ with $r$ degrees of freedom. When $\Gamma$ is unknown, $F_{T|Z}(t) = (1+\text{sign}(t)F_{F(r,\nu_n)}(t^2/r))/2$, where $F_{F(r,\nu_n)}$ is the distribution function of the $F$ distribution with $r$ and $\nu_n$ degrees of freedom.
\end{theorem}

\proof{Proof Sketch.}
A detailed proof of Theorem~\ref{thm:T|Z} based on the ``change of variables'' technique is provided in the online supplement. Here we give a short proof that provides richer insight into the result, but relies on a step that is essentially a consequence of the change of variables argument.
\begin{equation}\label{eqn:thm3sketch}
\begin{aligned}
F_{T|Z}(t) = P_n(T \le t | Z)
&= \begin{cases}
P_n(T \le 0 | Z) + P_n(0 \le T \le t | Z) , &\mbox{ when $t \ge 0$} \\
P_n(T \le 0 | Z) - P_n(0 \le T \le -t | Z) , &\mbox{ when $t < 0$}
\end{cases}\\
&= 1/2 + \text{sign}(t) P(||X||^2 \le t^2 | Z)/2, \text{ for } X = TZ\\
&= 1/2 + \text{sign}(t) P(||X||^2 \le t^2)/2.
\end{aligned}
\end{equation}

The last step in \refeq{eqn:thm3sketch} depends on the independence of $||X||^2$ and $Z$, as established in the proof in the online supplement. Intuitively, this result is a consequence of the structure of elliptical distributions, as discussed in, e.g., \cite{joe2014}, in that such random vectors can be generated by independently generating the direction $X$, scaling by a square root of the scale matrix, selecting the distance $T$ along the scaled direction independently of $X$, and finally adding on the mean.
 
When $\Gamma$ is known, $X \sim N(0,I)$, so $||X||^2 \sim \chi^2_r$. 

When $\Gamma$ is unknown, $X \sim t_{\nu_n}(0,I) = N / \sqrt{Y/\nu_n}$ for independent $N \sim N(0,I)$ and $Y \sim \chi^2_{\nu_n}$. Therefore $$||X||^2 = \frac{N^T N}{Y/\nu_n}$$ where $N^T N \sim \chi_r^2$, so $||X||^2/r \sim F(r,\nu_n)$.\Halmos
\endproof 

To find $\ti(Z)$ and $\ta(Z)$, we can solve linear programs with objectives minimizing or maximizing $t$, with decision variables $t \in \mbR, \bm{a} \in \mbR^{r \times d}, \bm{b} \in \mbR^{d}$, and the constraints (\ref{LS:big}), replacing $g(\xv)$ by $\bm{\mu} + (\Lambda^{1/2}Z)t$:
\begin{equation} \label{LP:big}
\begin{aligned}
& \ti = \displaystyle \text{min } t \hspace{1em} (\ta = \displaystyle \text{max } t) \\
\text{s.t.} \hspace{1em} & \bm{a}^T \xv + \bm{b} = \bm{\mu} + (\Lambda^{1/2}Z)t \;\\
& \bm{a}_i^T \xv_j + \bm{b}_i \leq \bm{\mu_j} + (\Lambda^{1/2}Z)_j t, \;\mbox{ for all } i \in \{1,\ldots,r\} \mbox{ and } j\neq i \mbox{, }  j \in \{1,\ldots,r\}
\end{aligned} \tag{LP}
\end{equation}

The linear program \ref{LP:big} can be decomposed into $r$ smaller LP's, with constraints \ref{LS:infeasDecomp} and variables $t \in \mbR, \bm{a}_i \in \mbR^{r}, b_i \in \mbR$:
\begin{equation} \label{LP:decomp}
\begin{aligned}
& \ti(i) = \text{min } t \hspace{1em} (\ta(i) = \text{max } t) \\
\text{s.t.} \hspace{1em} & \bm{a}_i^T \xv_i + b_i = \bm{\mu} + (\Lambda^{1/2}Z)t \;\\
& \bm{a}_i^T \xv_j + b_i \leq \bm{\mu_j} + (\Lambda^{1/2}Z)_j t, \; \mbox{ for all } j\neq i \mbox{, }  j \in \{1,\ldots,r\},
\end{aligned} \tag{LP(i)}
\end{equation}
and then
$$
\ti = \max_{i=1,2,\hdots,r} \ti(i), \quad \text{and} \quad
\ta = \displaystyle \min_{i=1,2,\hdots,r} \ta(i).
$$
This decomposition does not bring as much speed improvement as \ref{LS:infeasDecomp} does, because all the decomposed linear programs must be solved.

Now we have all the pieces needed for the conditional Monte Carlo method.
\begin{algorithm}[H]
\caption{A conditional Monte Carlo estimator $\widetilde{p}_n$ for $p_n = P(\f \in \mbC | \mathscr{A}_n)$.}
\begin{algorithmic}[1]\label{alg:pconvex2}
\REQUIRE Posterior distribution of $\f|\mcA_n$ obtained from Algorithm \ref{alg:main} with mean $\bm{\mu_n}$ and covariance $\Lambda_n$; Number of Monte Carlo samples $m$ needed
\FOR{$k = 1,\hdots,m$}
\STATE Uniformly generate a vector $z_k$ on the surface of a unit sphere (by generating a standard Gaussian and normalizing it to a unit vector).
\STATE Determine integration boundaries $\ti(z_k)$ and $\ta(z_k)$.
\STATE Set $\widetilde{P}_n(k) = F_{T|Z}(\ta(z_k)) - F_{T|Z}(\ti(z_k))$.
\ENDFOR
\STATE Calculate the mean $p_n^m$ and standard deviation $s_n^m$ of $(\widetilde{P}_n(k): k = 1, 2, \ldots, m)$. 
\RETURN $\widetilde{p}_n = p_n^m$ as an estimator of $P(\f \in \mbC | \mathscr{A}_n)$, along with the half-width $\tilde{h}_n = 1.96s_n^m / \sqrt{m}$ of a $95\%$ confidence interval.
\end{algorithmic}
\end{algorithm}

Relative to the other variance-reduction methods, conditional Monte Carlo takes much longer to produce an estimate in each iteration because it needs to solve two linear programs \ref{LP:big} and cannot ``skip" any of them as can be done when solving the decomposed feasibility problems \ref{LS:infeasDecomp}. 

\section{Numerical Results} \label{sec:numericalResults}
In this section we show numerical results on some test functions, assuming the more realistic case that the sampling variance is unknown. To select the $r$ sample points for a test function in $d$ dimensions, we first sample $d+1$ points uniformly at random within the (assumed compact) sample space $S$, and for each such random point, we generate a uniform random direction on the surface of the unit sphere. Each point-direction pair defines a line segment within $S$. Then we sample $3$ points uniformly at random on each line segment. This method generates $r = 3(d+1)$ sample points. We select the points to lie on line segments because doing so seems to improve the performance of the convexity test relative to just sampling points uniformly within $S$. We have used $3(d+1)$ sample points partly to keep the computation minimal, thereby enabling us to relatively easily explore the behavior of the algorithms and estimators in this section. In practice, one would likely use more points, though it is unclear exactly how many points should be chosen. The number of points is likely related to how certain one wishes to be about convexity or lack thereof. In each iteration of the sequential algorithm, we use $m=100$ Monte Carlo samples from the posterior predictive distribution to estimate a 95\% confidence interval for $p_n$. With each of our estimators, one can easily adjust the sample size $m$ to achieve a desired accuracy in the confidence interval widths of the estimators of $p_n$; using $m=100$ gave reasonable results in our experiments. We discuss the choice of $n$ in Section~\ref{sec:conclusion}.

\begin{figure}[h]
\FIGURE
{\includegraphics[scale=0.25]{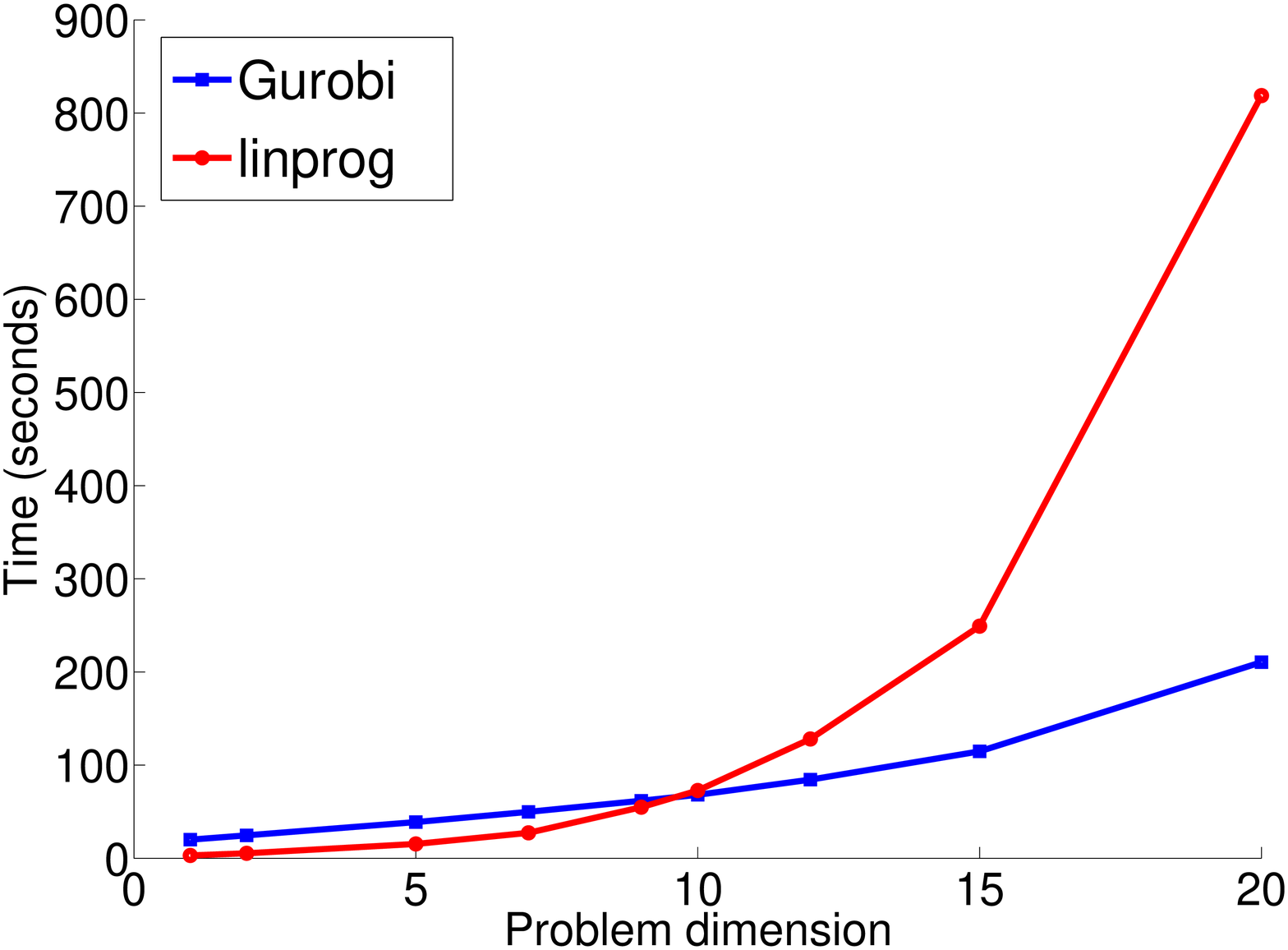}}
{The solving time vs. testing function dimension for the two linear programs in the conditional Monte Carlo method using {\tt Gurobi} and {\tt linprog}. {\tt Gurobi} is faster when the dimension $d$ exceeds 10, where $r = 33$ sample points are used. \label{fig:solver_time}}
{}
\end{figure}

Our procedure is implemented in Matlab and freely available in an online repository \citep{Jian2017}. The repository contains two versions. The first version uses only a standard Matlab installation, solving linear programs using the built-in {\tt linprog} function \citep{linprog}. The second version requires the installation of the packages {\tt CVX} \citep{cvx,gb08}, which is a package for specifying and solving convex programs, and {\tt Gurobi} \citep{gurobi}, a commercial optimization solver. We suggest the second version if a user has the requisite licenses, since {\tt Gurobi} seems more robust than {\tt linprog}. For example, we have found cases where {\tt linprog} was not able to find a feasible solution, whereas {\tt Gurobi} did. However, because of the overhead of {\tt CVX} in setting up the linear program in a format that {\tt Gurobi} is able to read, {\tt linprog} is usually faster when the problem dimension is low. When the problem dimension is high, the inefficiency of the interior-point-method used by {\tt linprog} outweighs the overhead of {\tt CVX}. Figure \ref{fig:solver_time} compares solving times in seconds by these two solvers for the linear programs \refeq{LP:big}, tested with the sample function $f(\xv) = ||\xv||^2, \xv \in [-1,1]^{d}$ for different values of the problem dimension $d$. 

All test cases are run on a desktop with a 4-core Intel Core i7-3770 3.40 GHz processor with 16G memory, running Matlab R2013a on 64-bit Windows 7.

\subsection{A Strictly Convex Function}
We use $f(\xv) = ||\xv||^2$ in this section as the test function.

First, we compare vanilla Monte Carlo with the variance reduction methods in Section~\ref{sec:varianceReduction}, showing 95\% confidence intervals for the estimated probability of convexity, the time per iteration, and the efficiency per iteration. Here the efficiency of the Monte Carlo estimator $\tilde{p}_n$ is defined as the inverse of the product of the computational time per replication and the variance of one replication; see, e.g., \cite{Glynn1992}. We first take the dimension $d = 1$, on the sample space $[-1,1]$. The sampling covariance matrix has equal constant variances of $0.01$ on the diagonal, and we use a Gaussian kernel of $10^{-4}\exp\{-||\xv_i - \xv_j||^2 / 2\}$ for the off-diagonal components. Hence the noise at different points is positively correlated, and the correlation is stronger between closer points \citep{Rasmussen2005}.

We use {\tt linprog} instead of {\tt Gurobi} to avoid the time overhead incurred by {\tt CVX}, since the dimension $d=1$. For the change of measure method, a new set of samples is obtained every iteration for the first 30 iterations, and every 5 iterations thereafter. For the acceptance-rejection method we start to reuse samples only after the first 30 iterations. Thus the first 30 iterations of these two methods are exactly the same as vanilla Monte Carlo.

\begin{figure}[htb]
\FIGURE
{\begin{tabular}{ c c c }
\includegraphics[width=54mm]{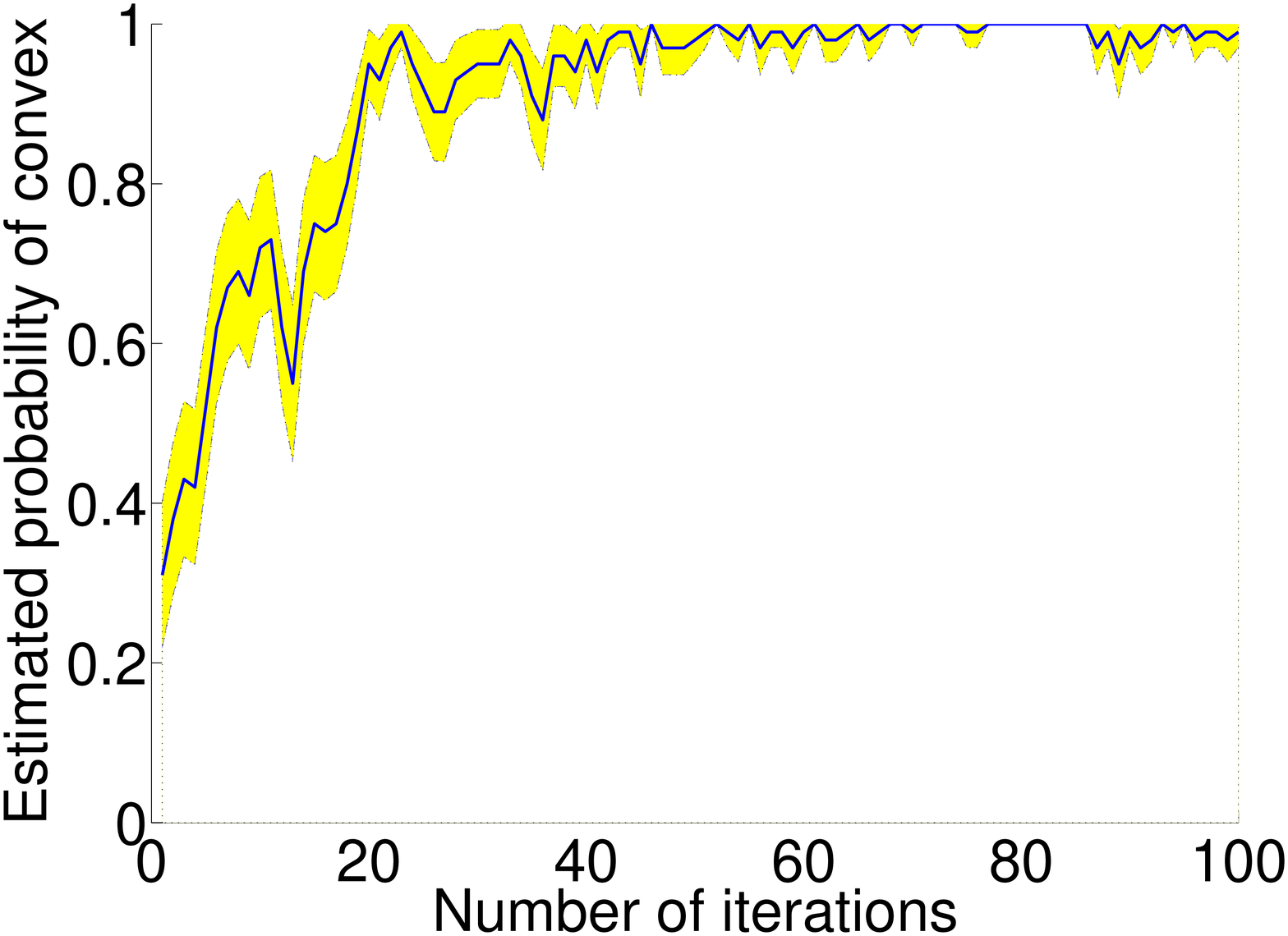} &  \includegraphics[width=54mm]{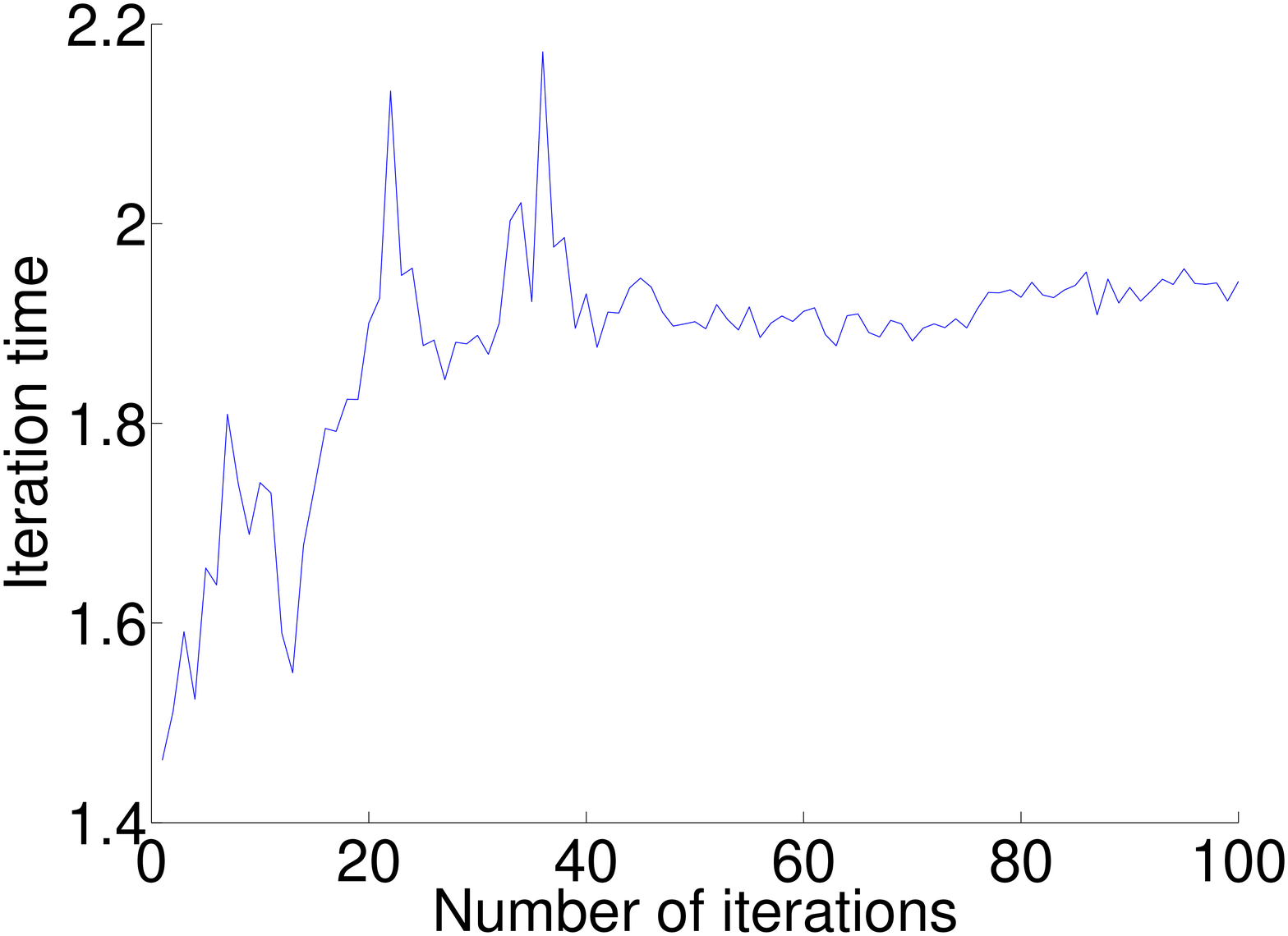} &  \includegraphics[width=54mm]{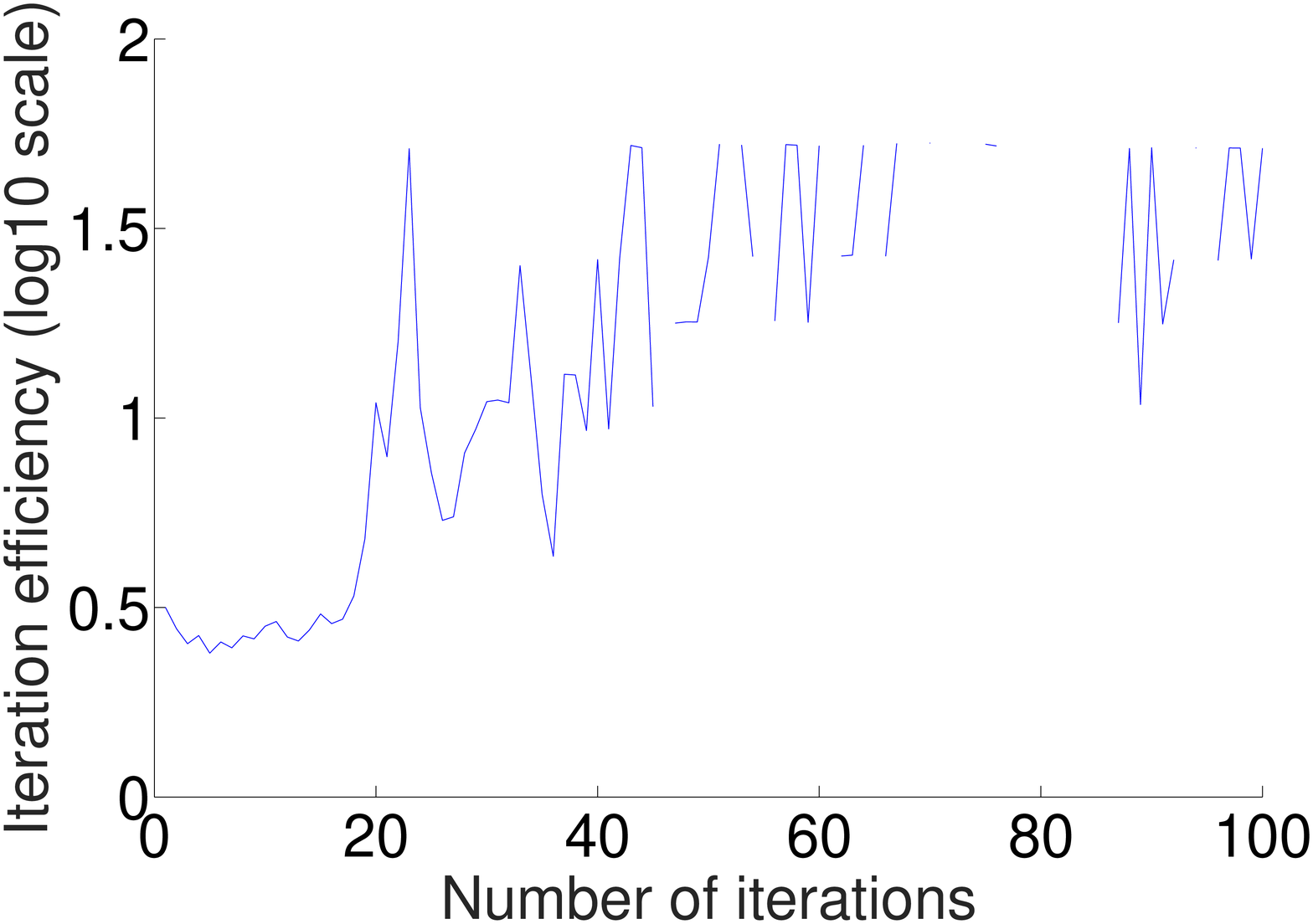} \\
\includegraphics[width=54mm]{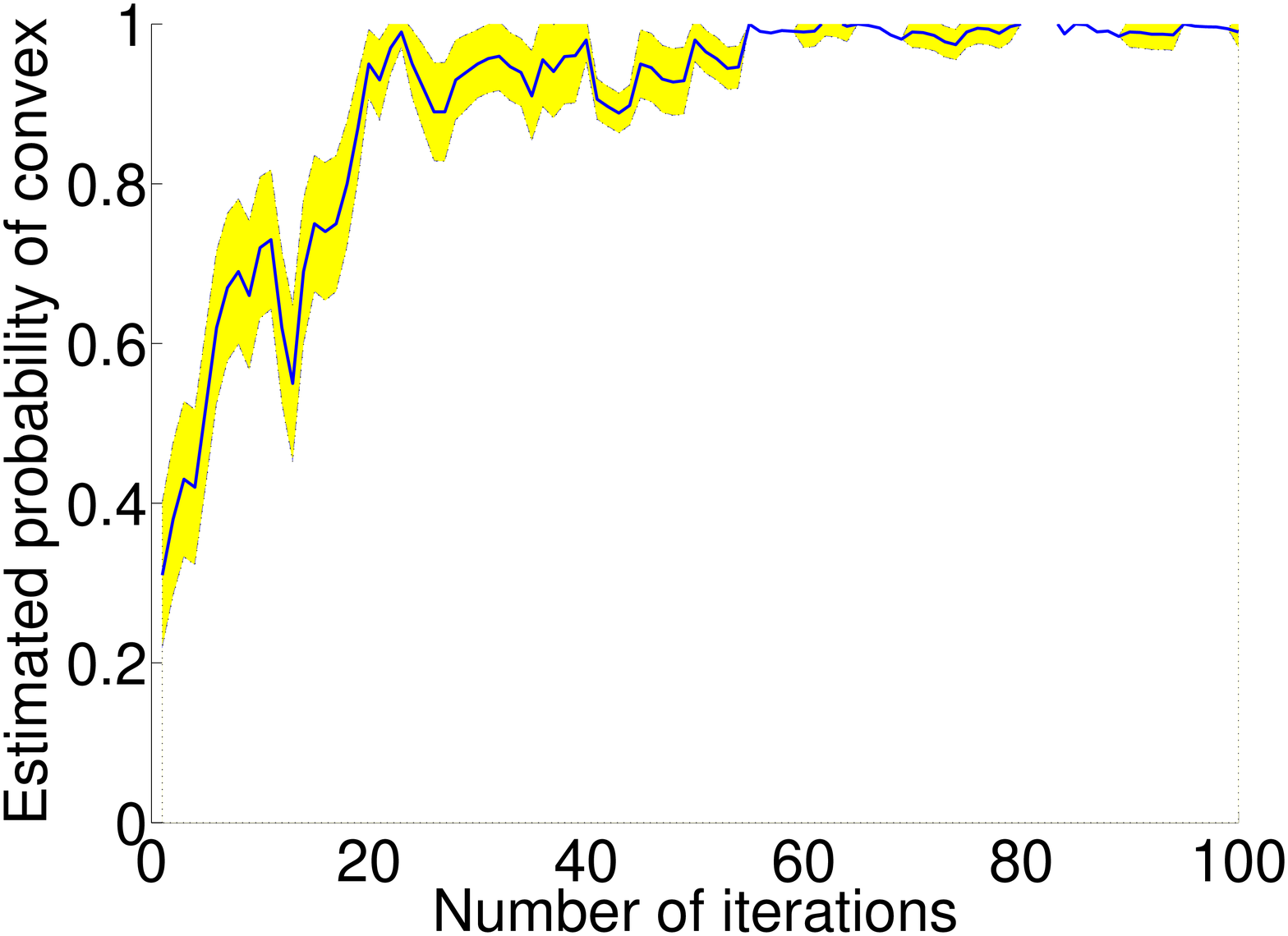} & \includegraphics[width=54mm]{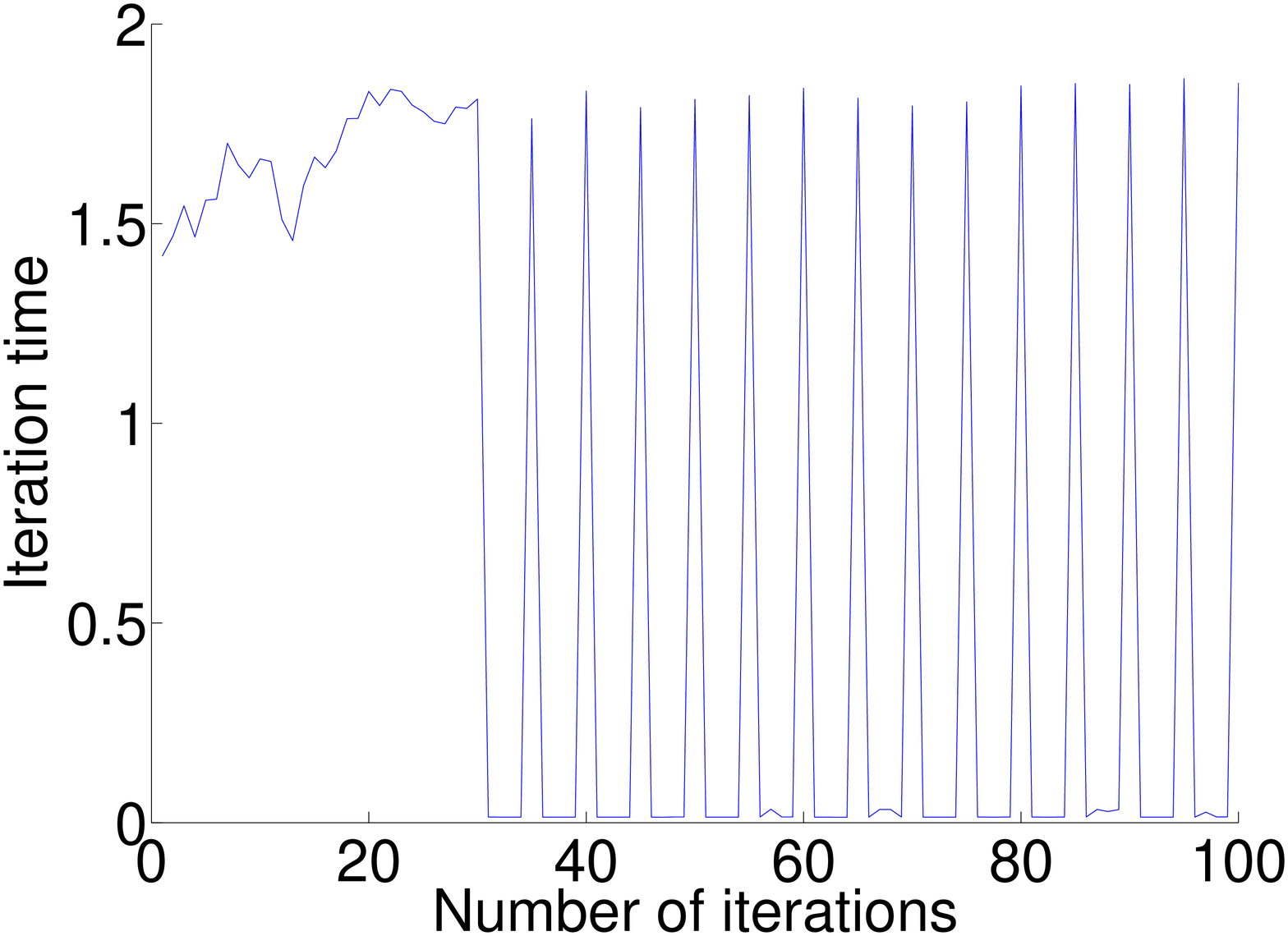} & \includegraphics[width=54mm]{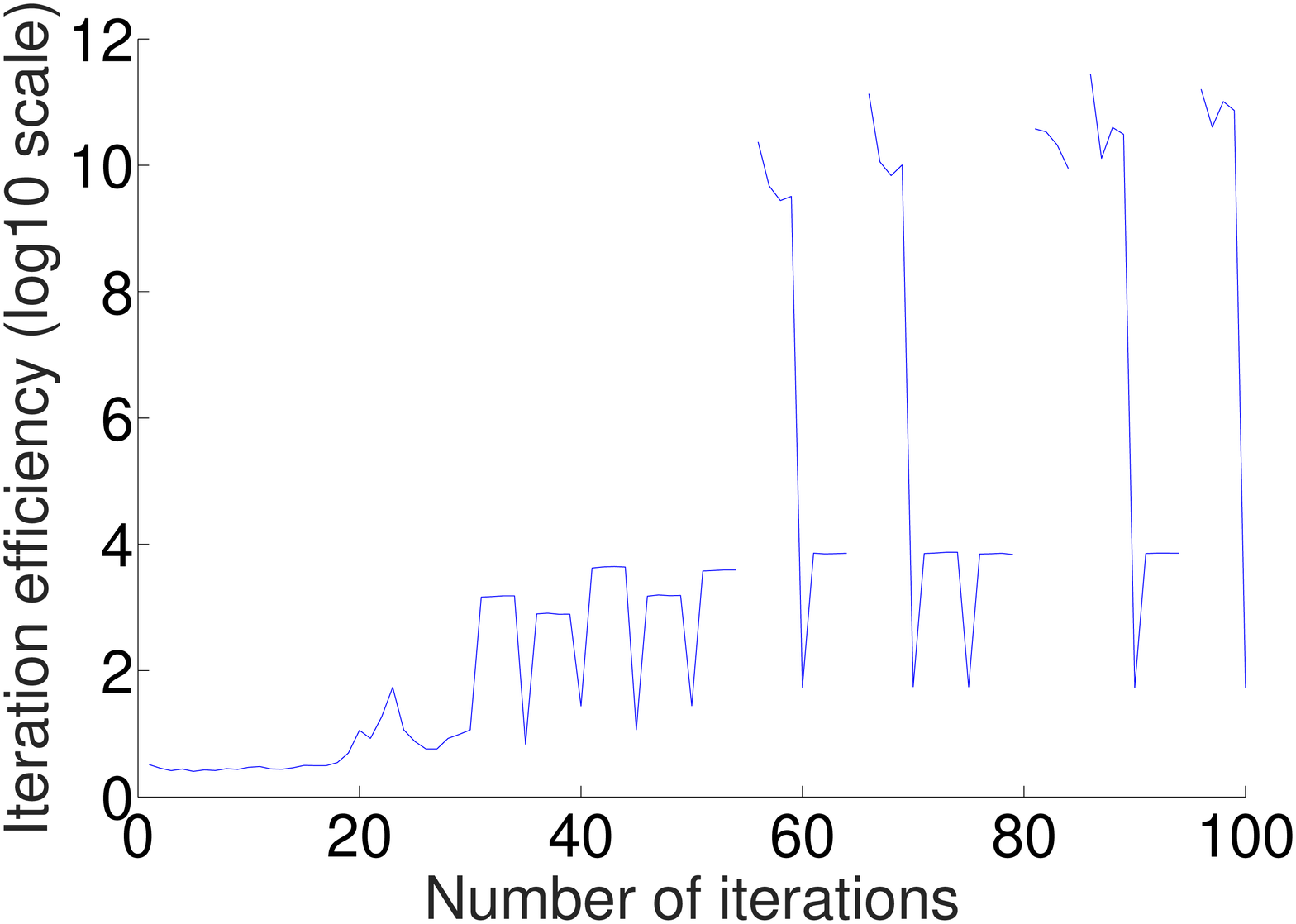} \\
\includegraphics[width=54mm]{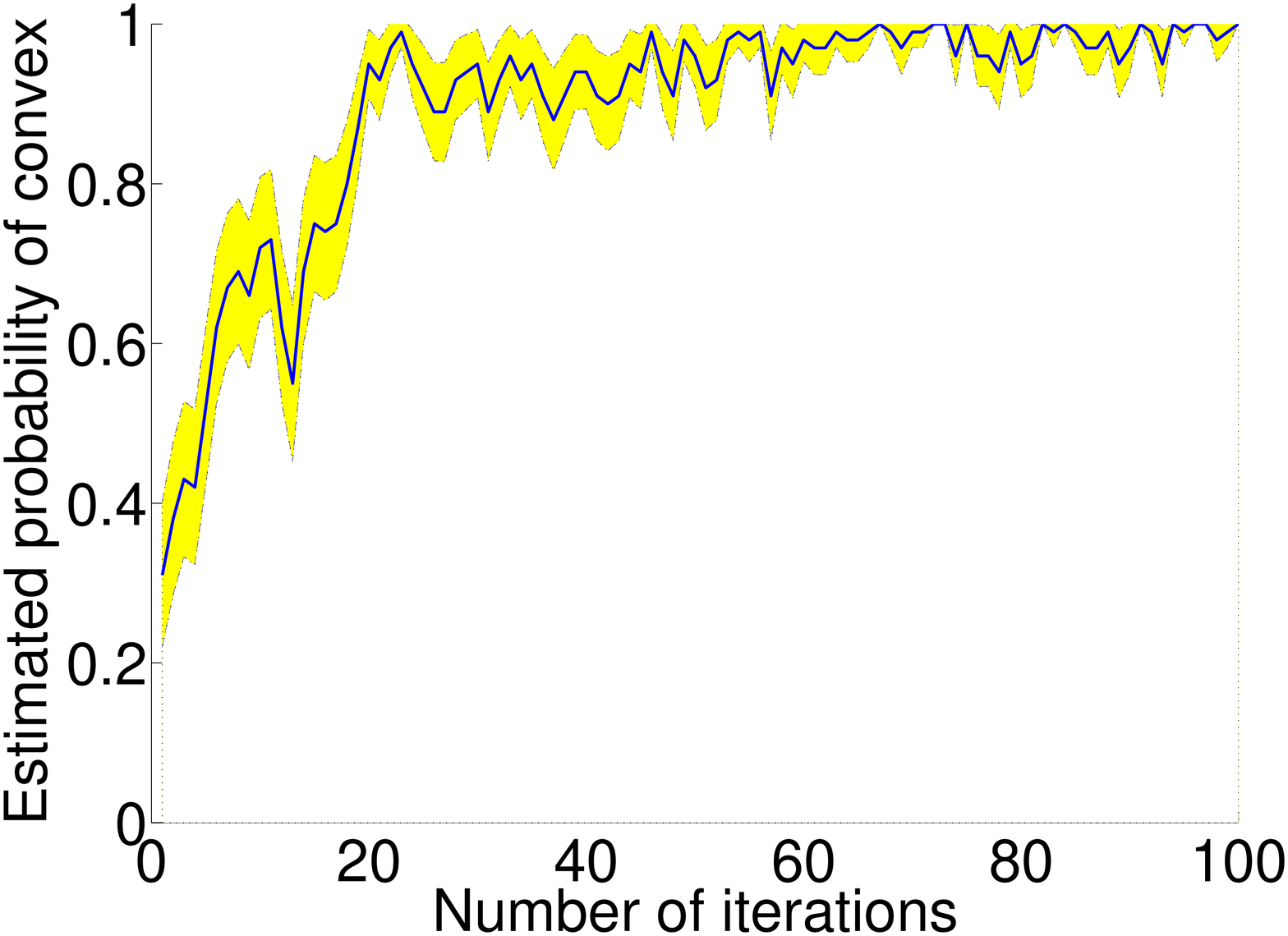} & \includegraphics[width=54mm]{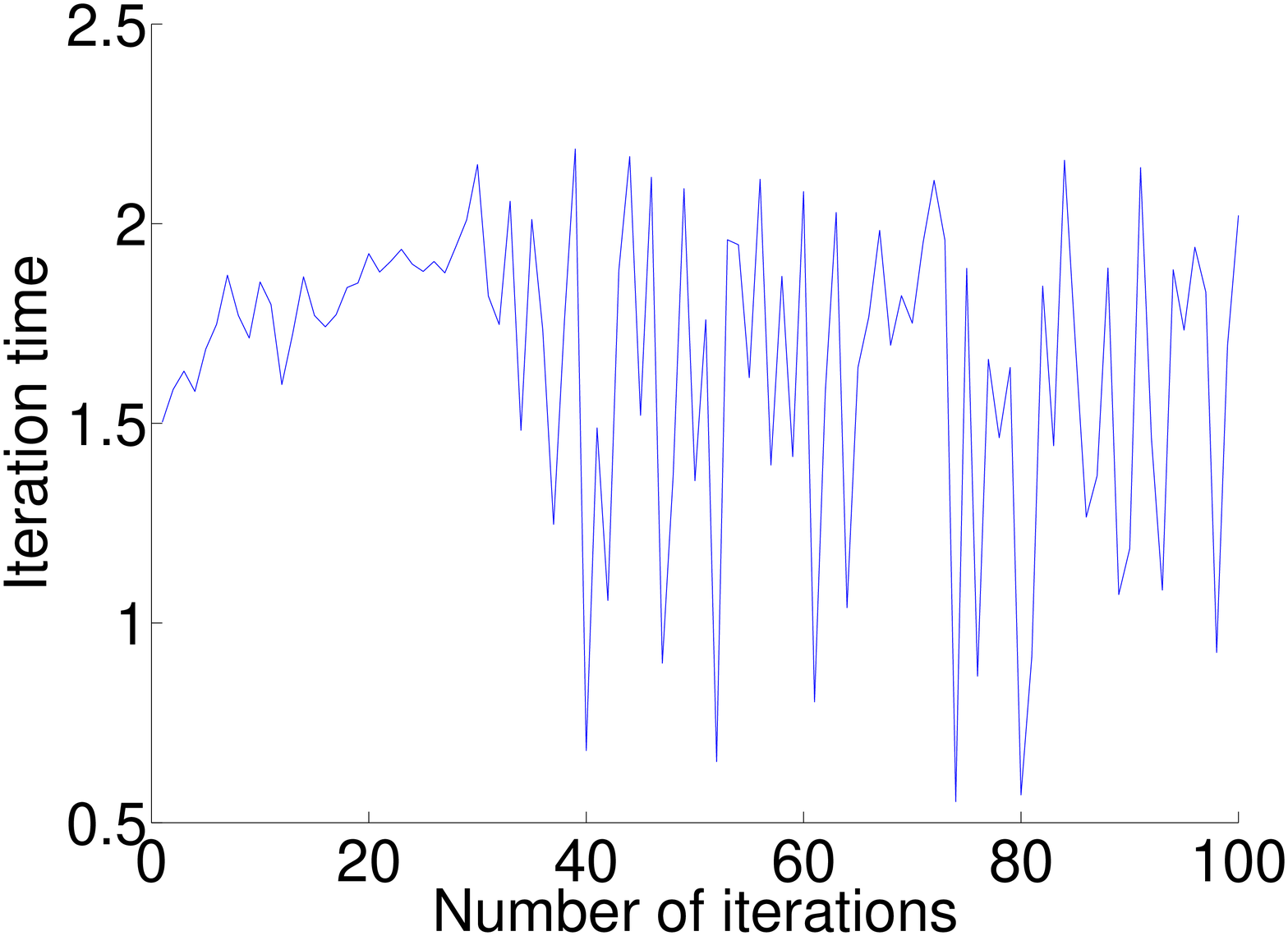} & \includegraphics[width=54mm]{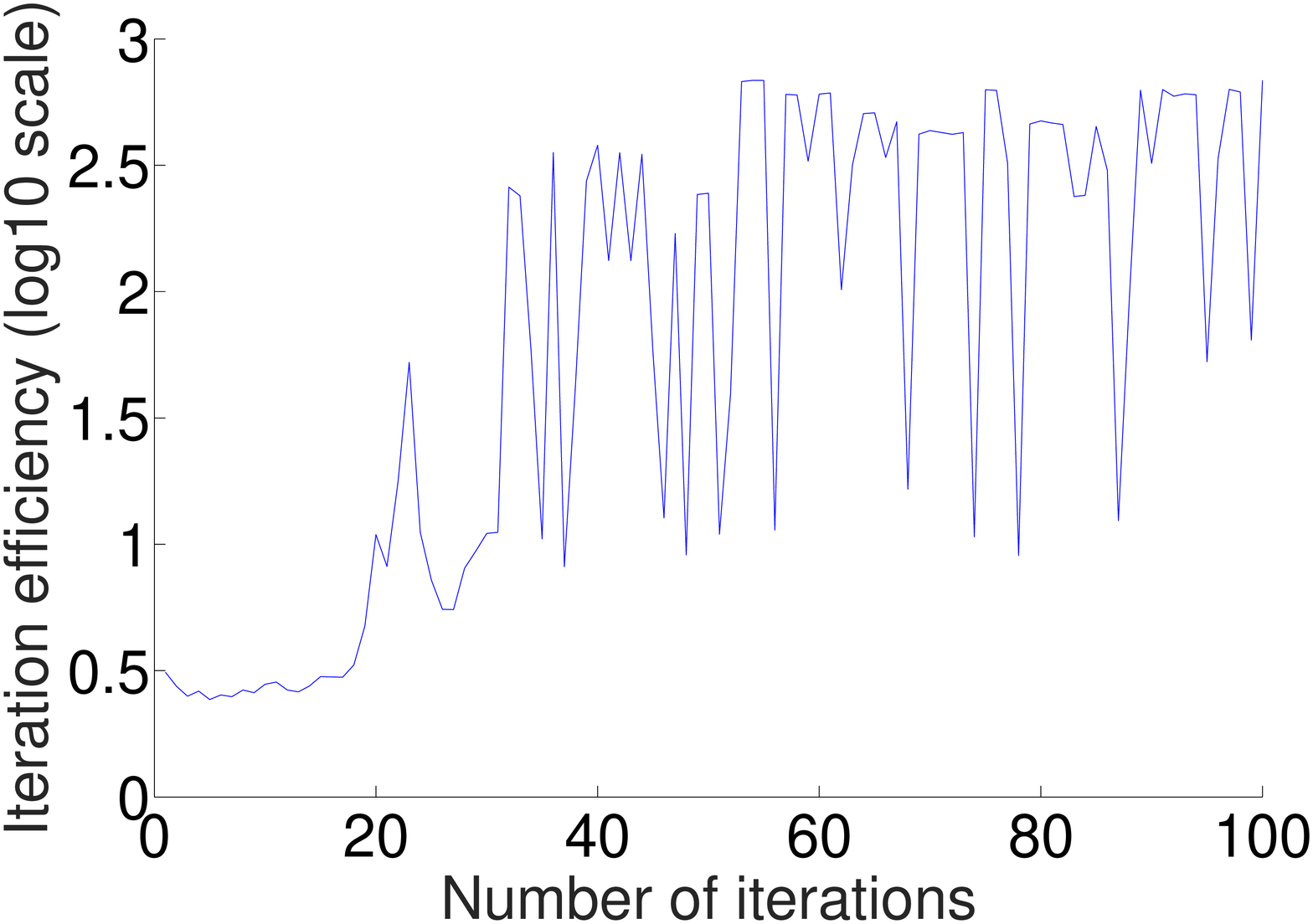} \\
\includegraphics[width=54mm]{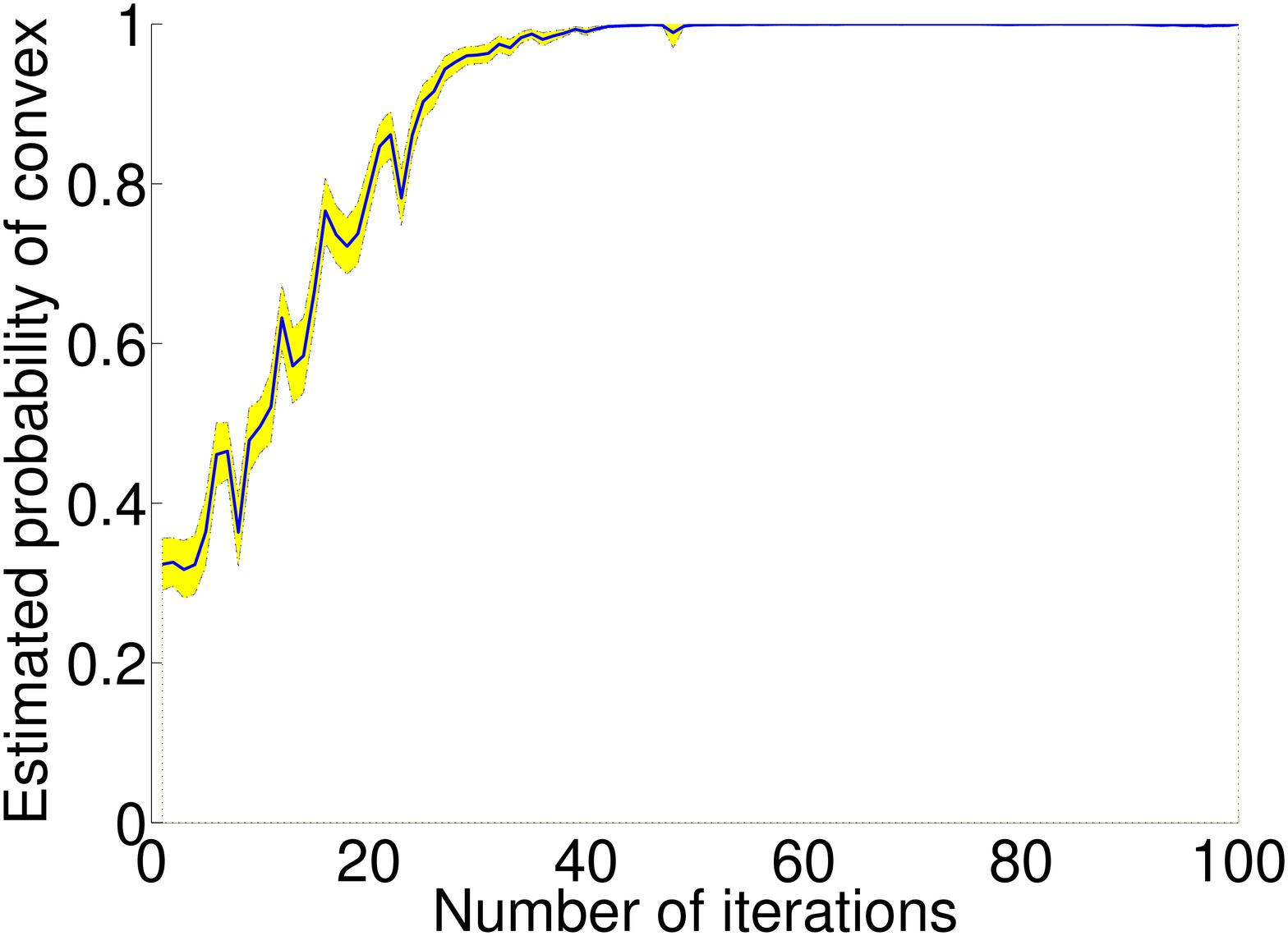} & \includegraphics[width=54mm]{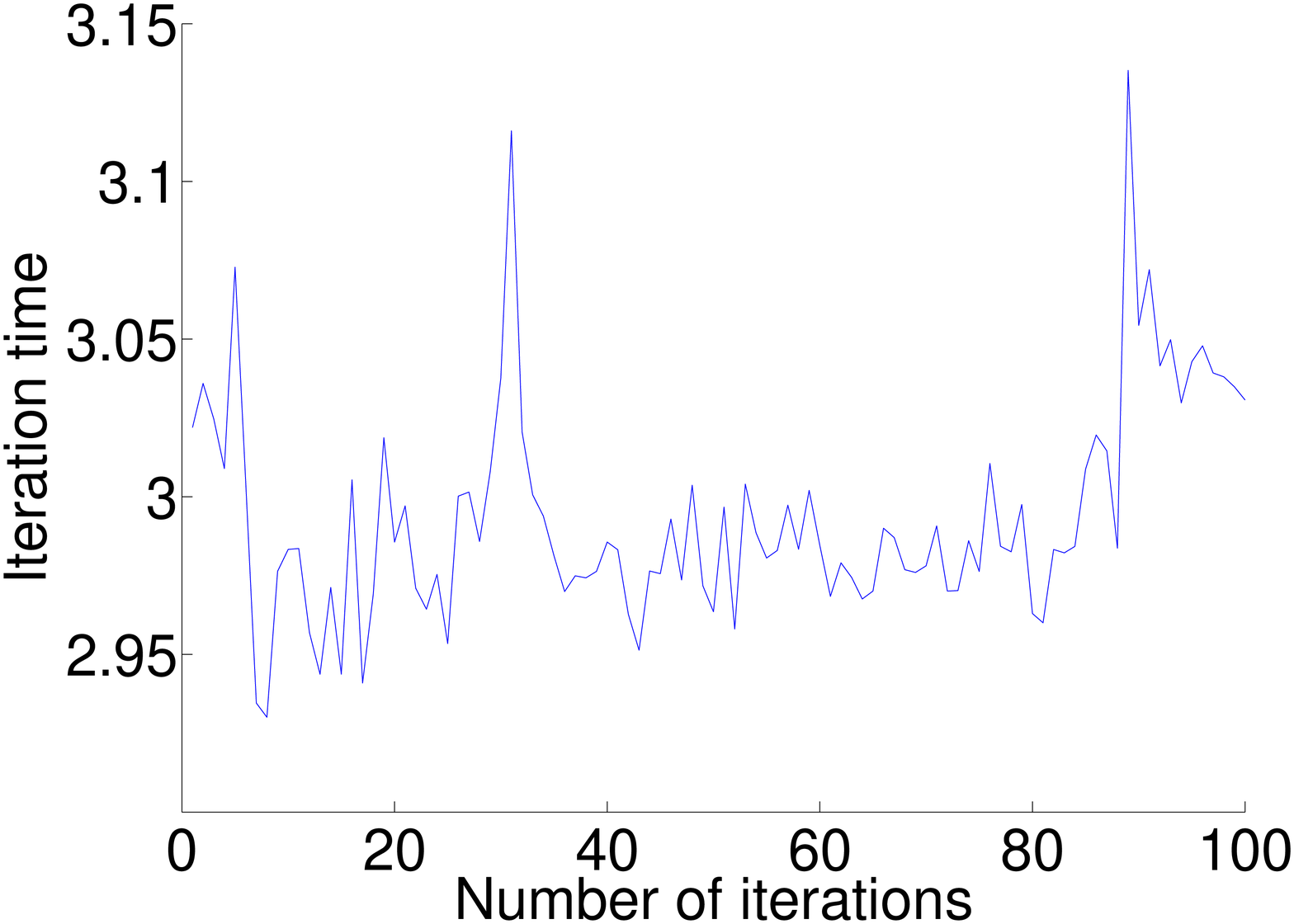} & \includegraphics[width=54mm]{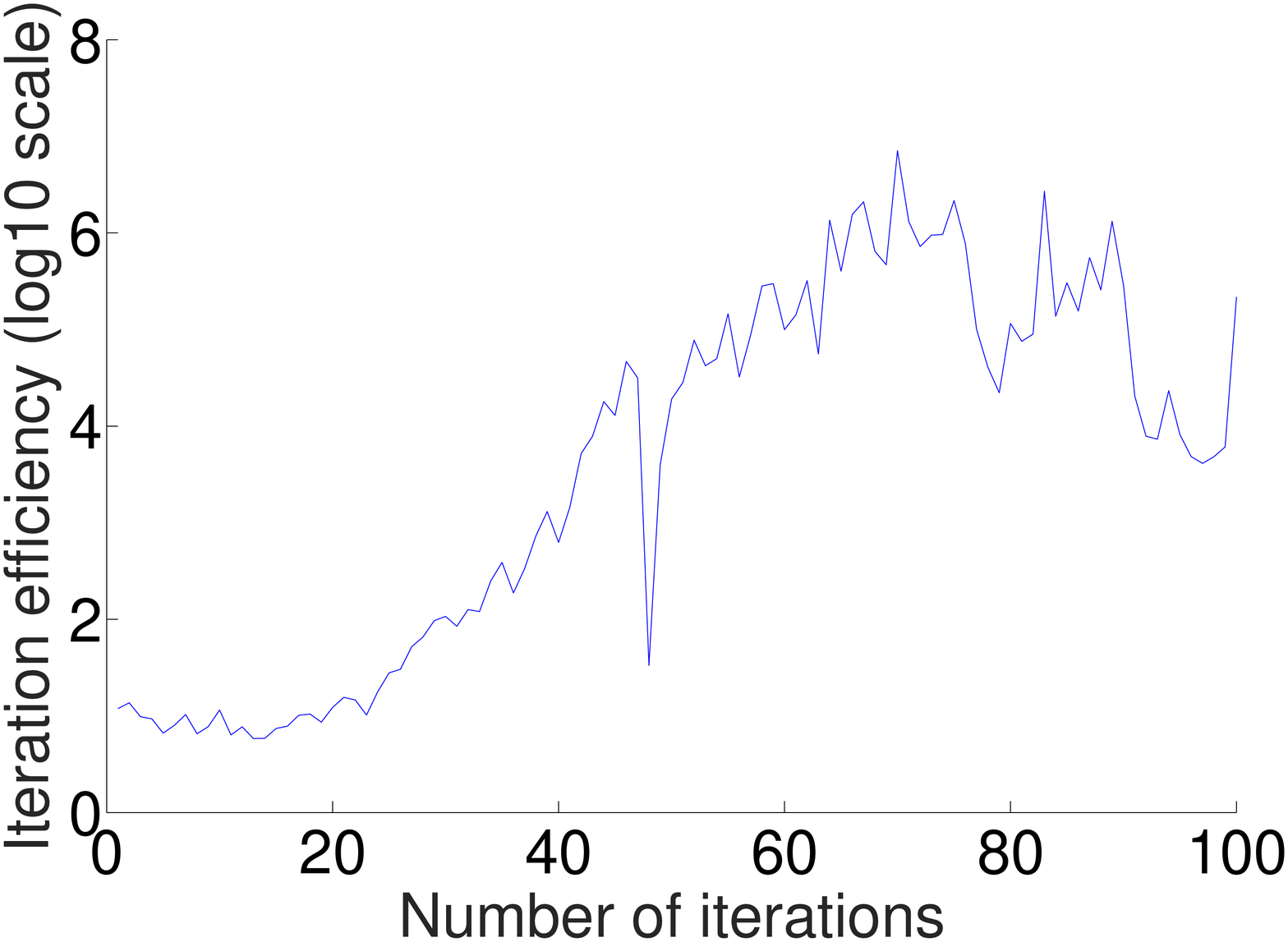}
\end{tabular}
}
{Comparisons of the estimated probability of convexity, the iteration time (in seconds), and the log (base 10) efficiency (left to right) of vanilla Monte Carlo, change of measure, acceptance-rejection, and conditional-Monte-Carlo (top to bottom) methods applied to a one-dimensional strictly convex function.\label{fig:convex_1d}}
{}
\end{figure}

Figure~\ref{fig:convex_1d} shows that the estimated probability of convexity increases to 1 for all methods. Later in the iterations, the change of measure method can return greater-than-one estimates due to the poor behavior of the likelihood ratio as discussed in Section~\ref{sec:changeofmeasure}. Among all methods, the conditional Monte Carlo method has the smallest variance but takes the longest time to compute. (This difference in the computational time becomes more significant for higher-dimensional test functions when we later experiment on a 30-dimensional function.) For the 1-dimensional convex function here, taking both computational time and variance into consideration, we observe that conditional Monte Carlo has the highest overall efficiency. The efficiency of vanilla Monte Carlo is the lowest. The efficiency plots occasionally break when the sample variance of the estimator is 0, where all the linear systems \refeq{LS:big} are feasible. This also happens for the change of measure method because that method corresponds with vanilla Monte Carlo every 5 iterations after the first 30. The efficiency of the change of measure method and the acceptance-rejection method both increase whenever they reuse the samples from a previous iteration because the linear feasibility problems need not be solved. The change of measure method occasionally has a very large efficiency because of the small sample variance of the estimate. This happens in later iterations when the posterior density is very concentrated. In this case all the Monte Carlo samples are close to the mean, giving almost identical posterior densities and similar likelihood ratios. When all reused samples are convex (corresponding to the iterations where the vanilla Monte Carlo method has infinite estimated efficiency), the change of measure estimator has almost 0 variance. However, due to the heavy tail behavior of the likelihood ratio, it is risky to trust the change of measure estimator values, as we see when the change of measure method estimates a probability greater than 1. The acceptance-rejection estimator has slightly lower estimated efficiency, but the estimator is more trustworthy in that it is statistically identical to vanilla Monte Carlo.

\begin{figure}[htb]
\FIGURE
{
\centering
\begin{tabular}{c c c}
\includegraphics[width=54mm]{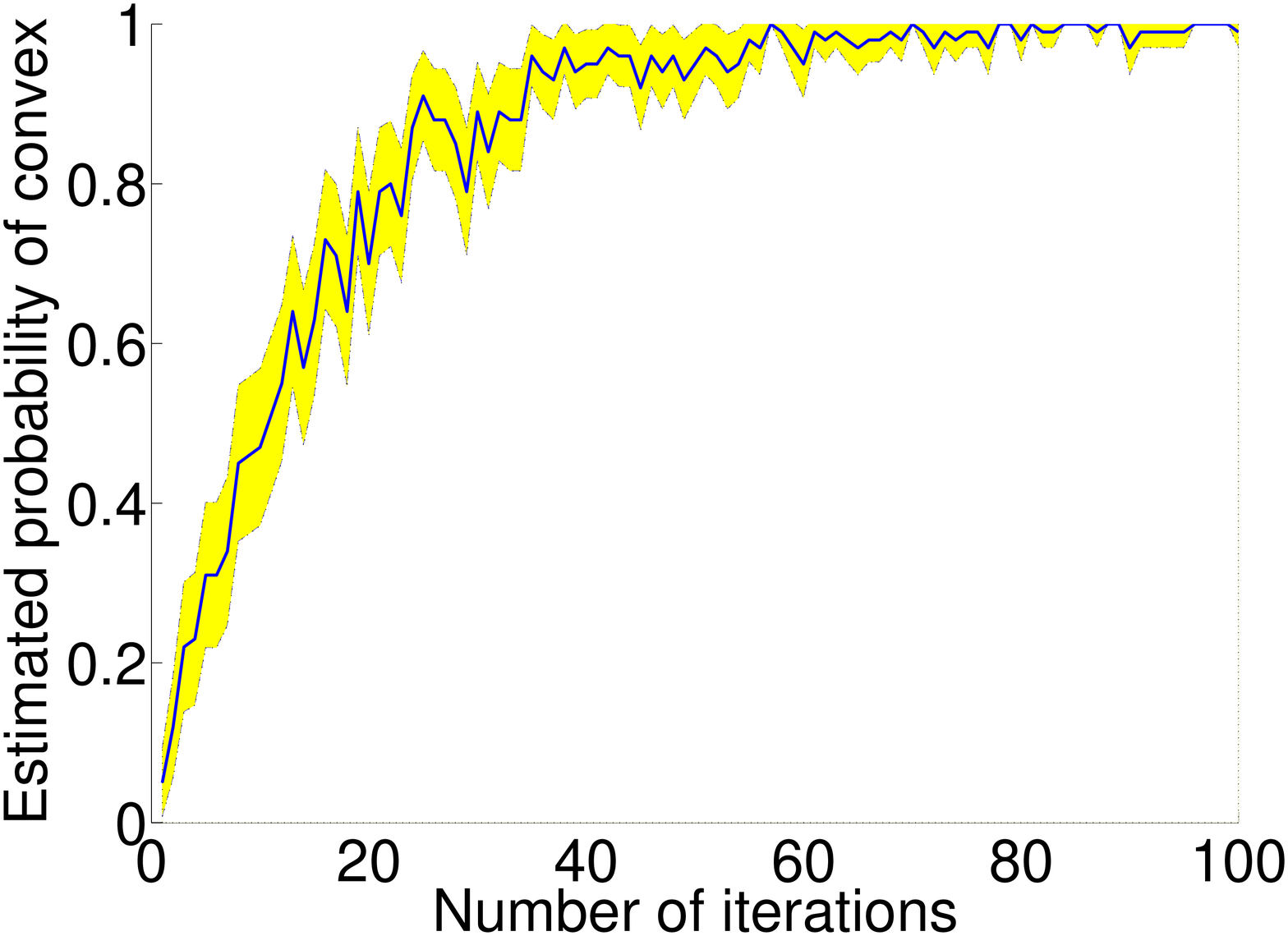} &
\includegraphics[width=54mm]{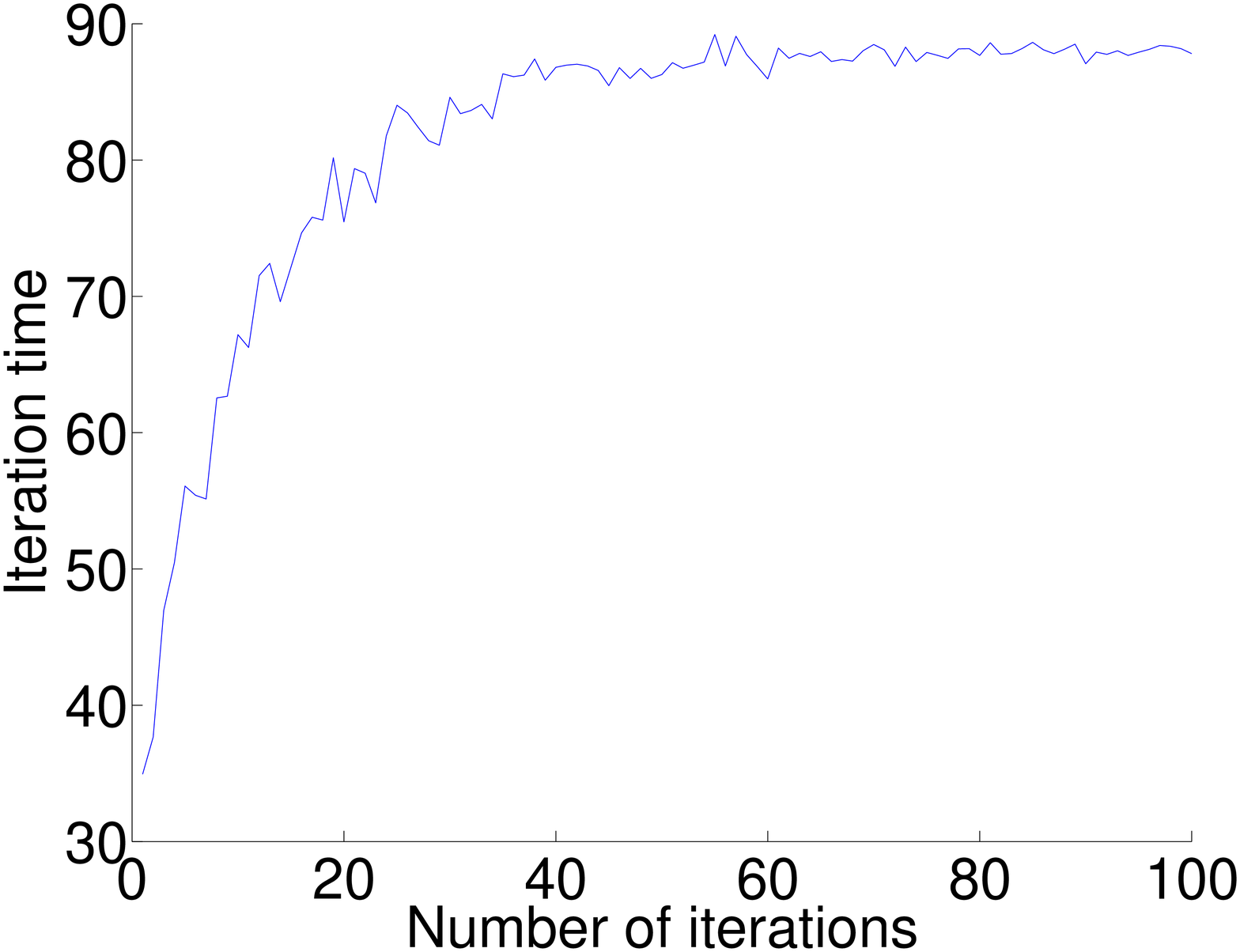} &
\includegraphics[width=54mm]{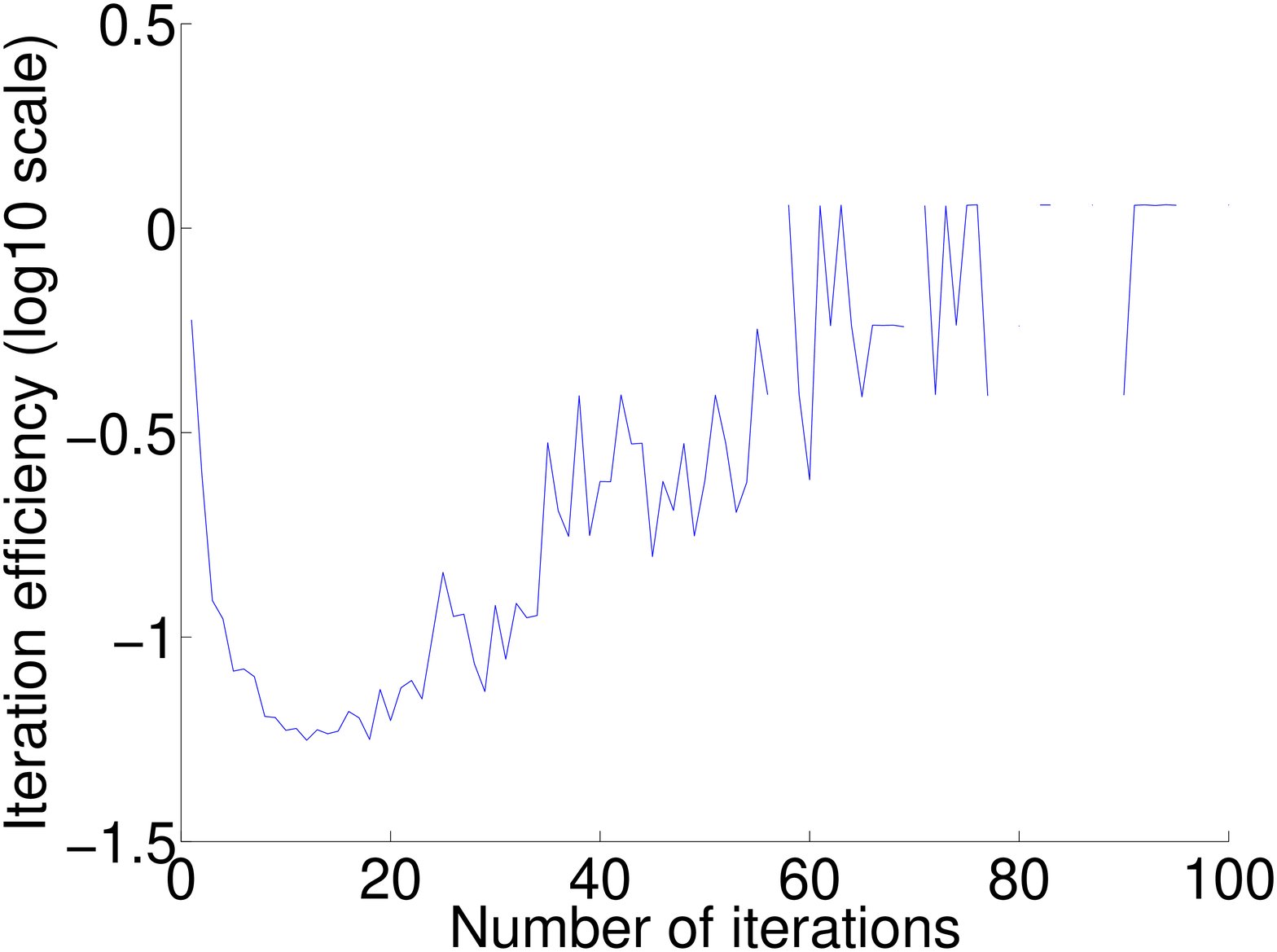}\\
\includegraphics[width=54mm]{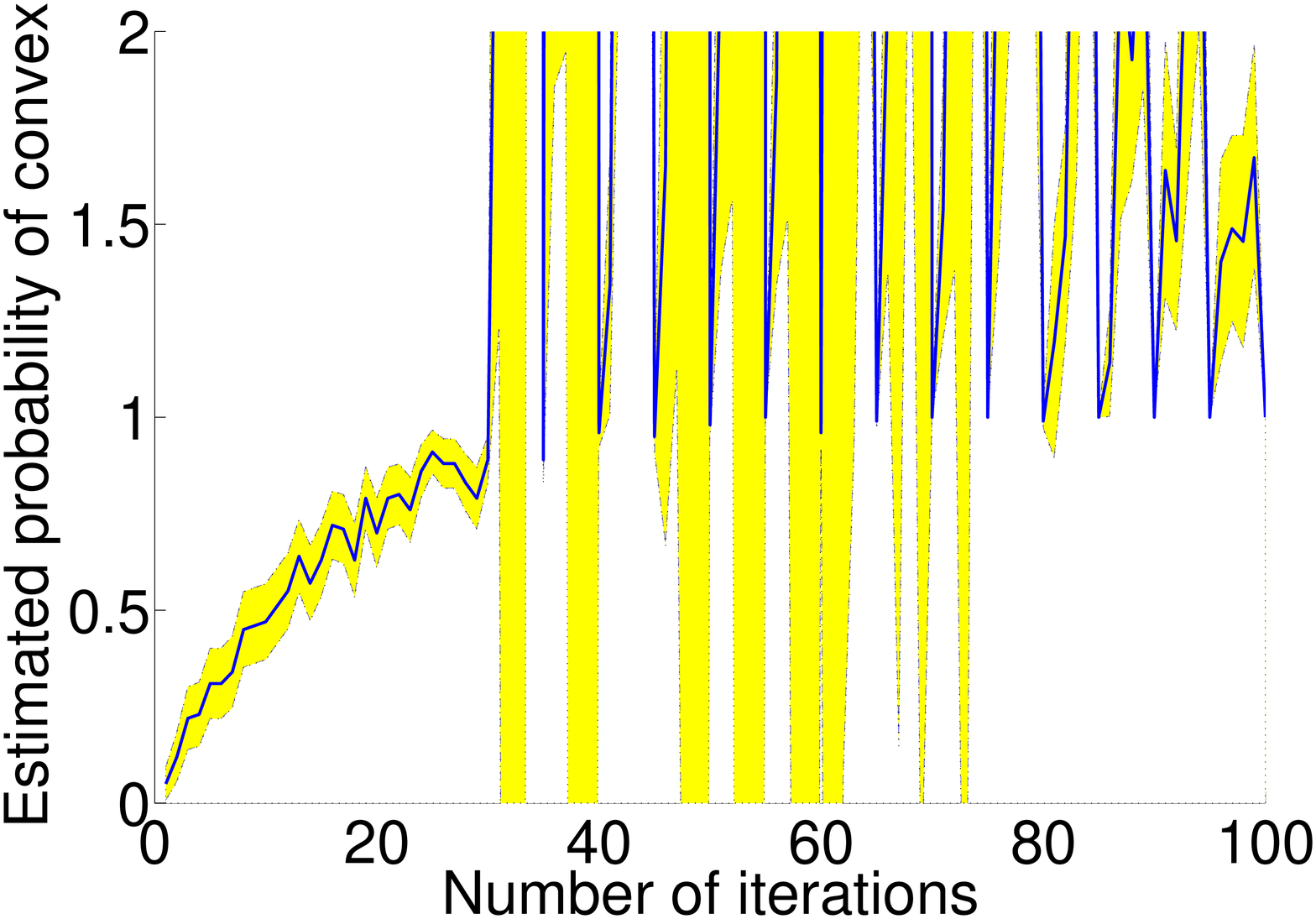} &
\includegraphics[width=54mm]{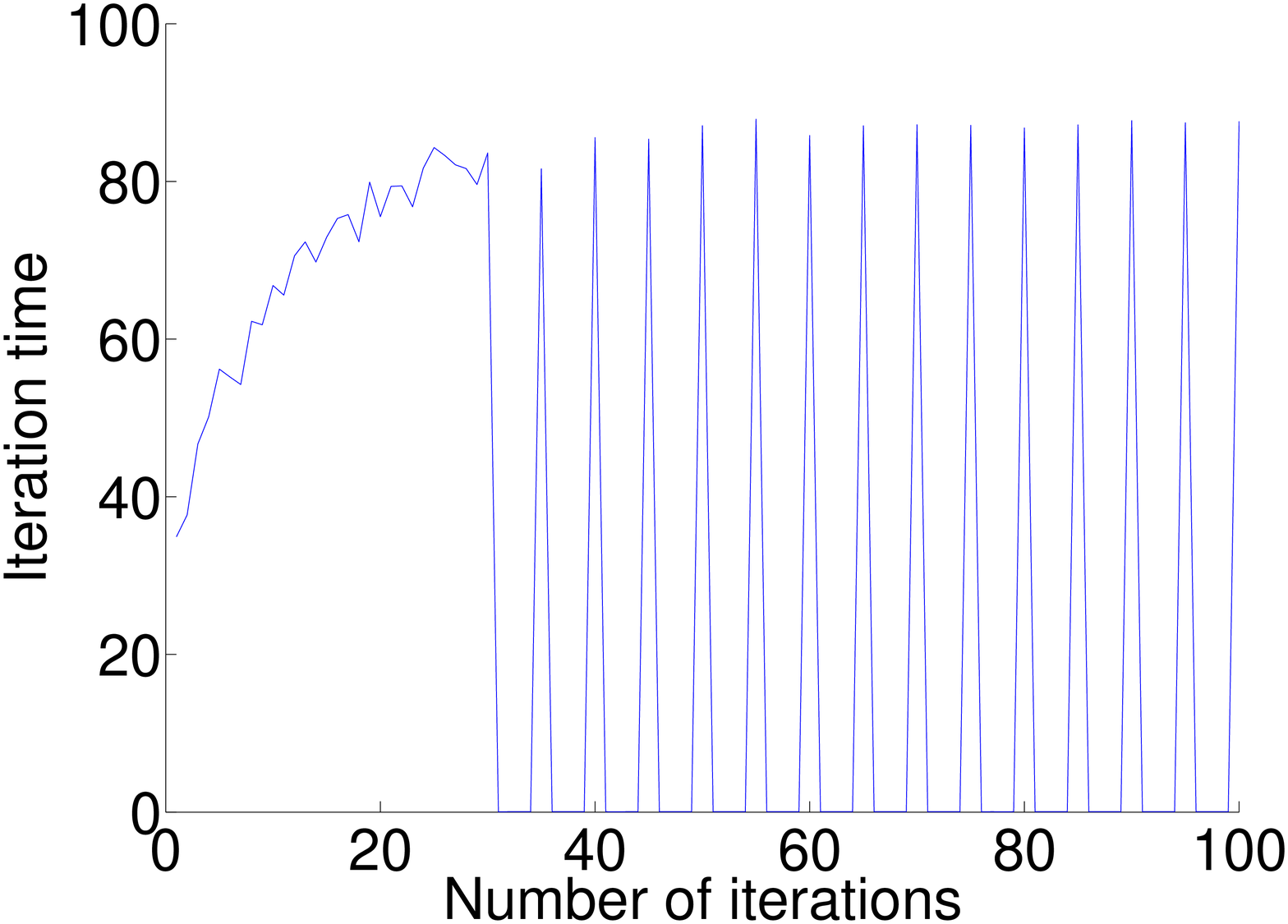} &
\includegraphics[width=54mm]{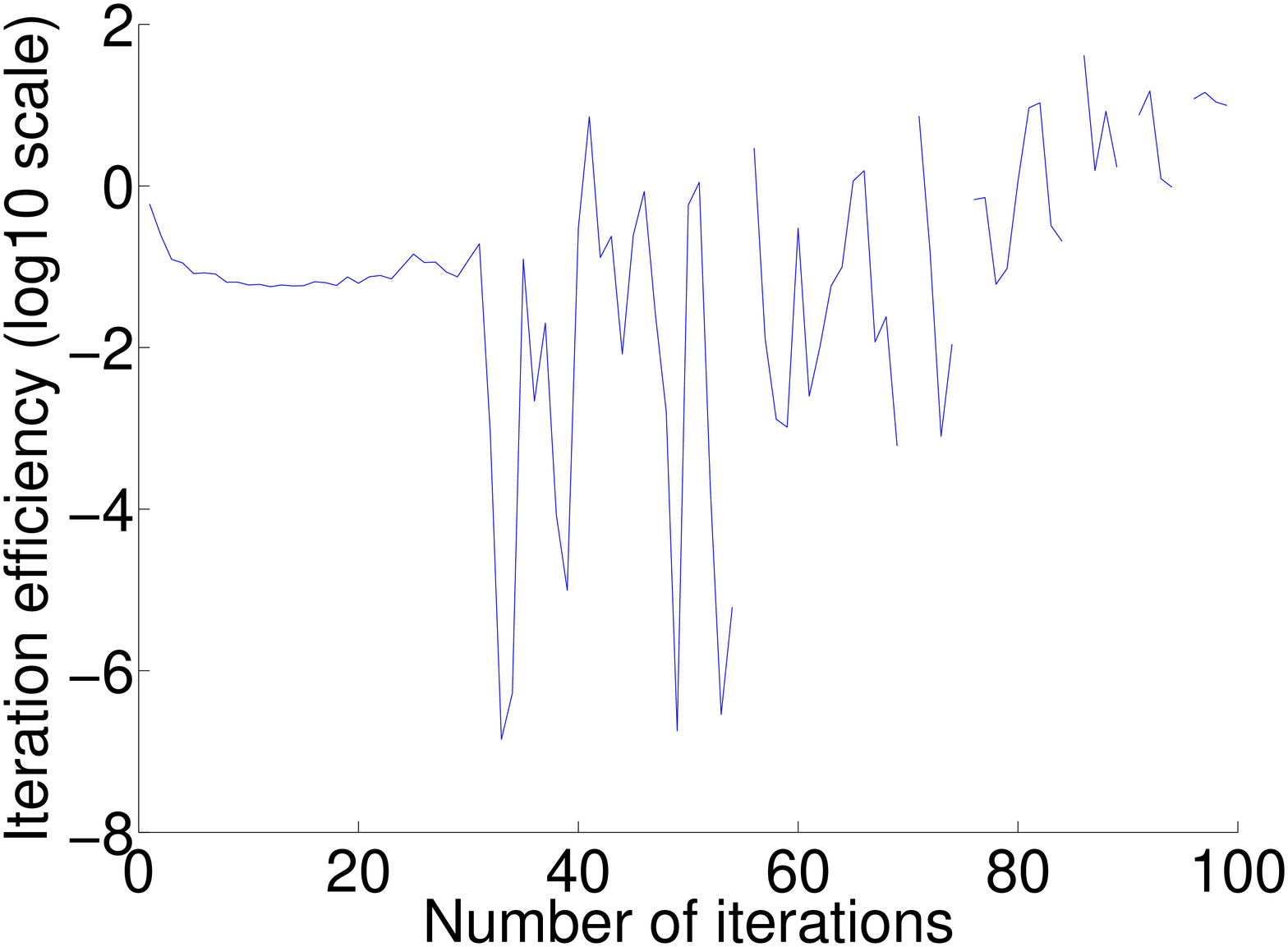}\\
\includegraphics[width=54mm]{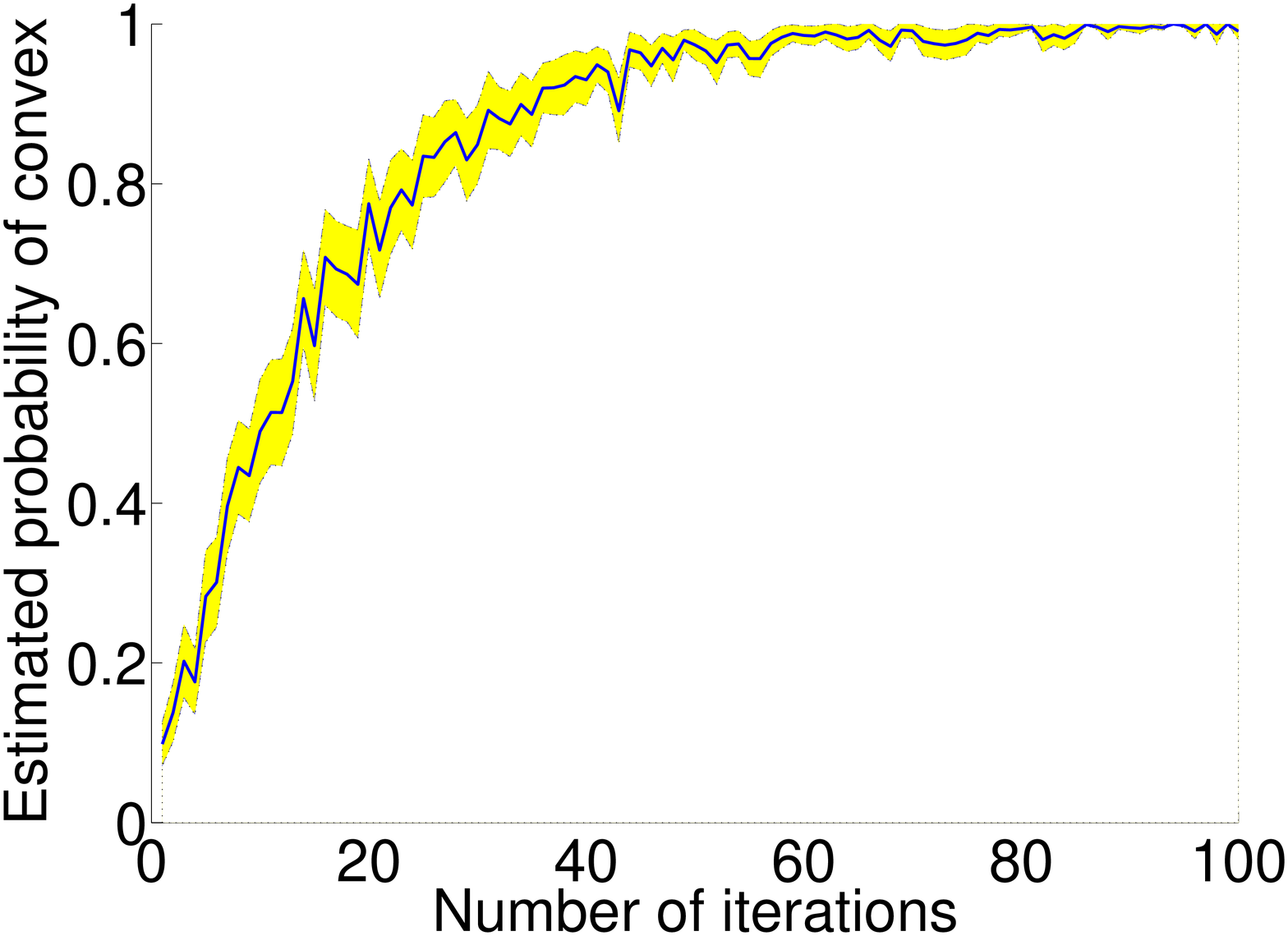} &
\includegraphics[width=54mm]{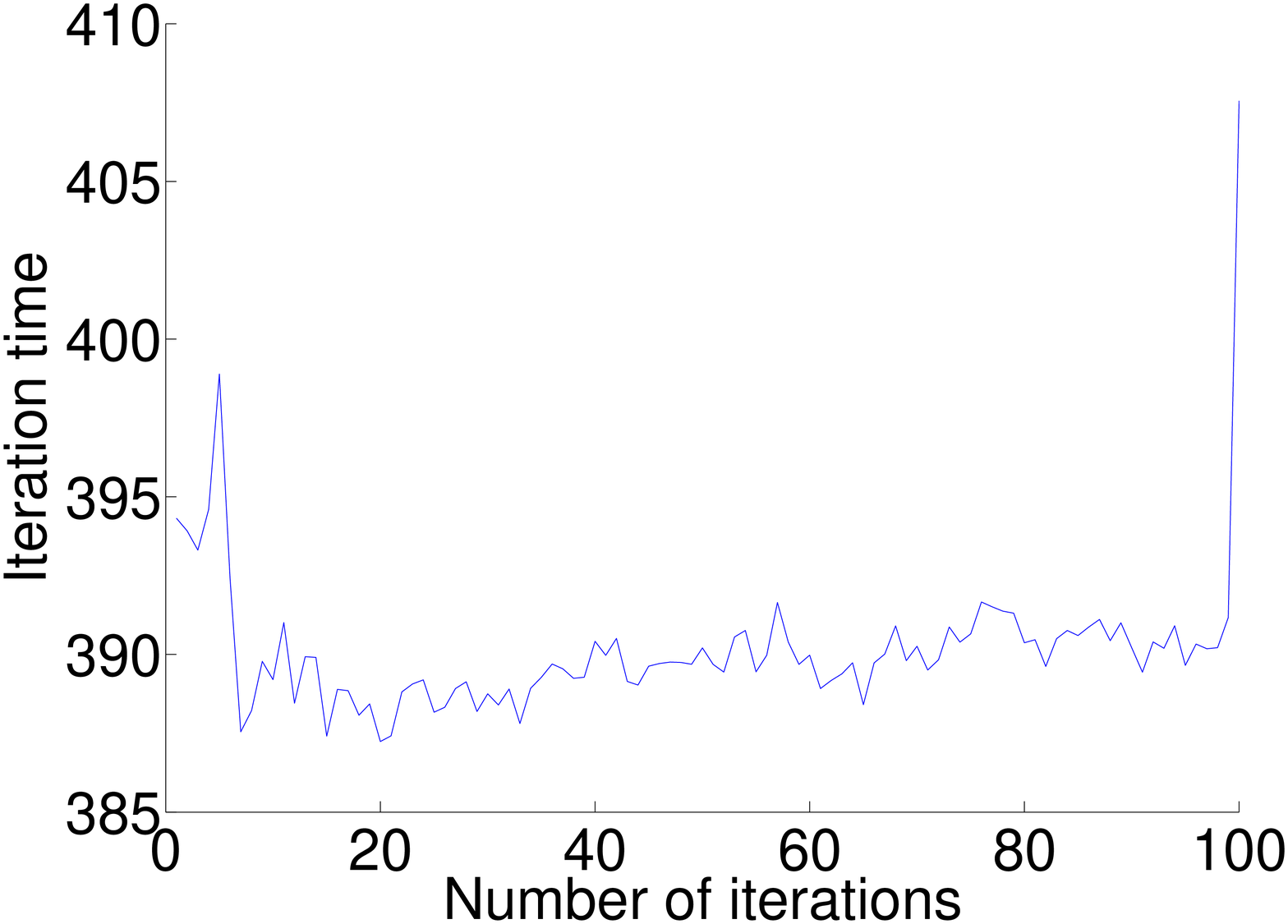} &
\includegraphics[width=54mm]{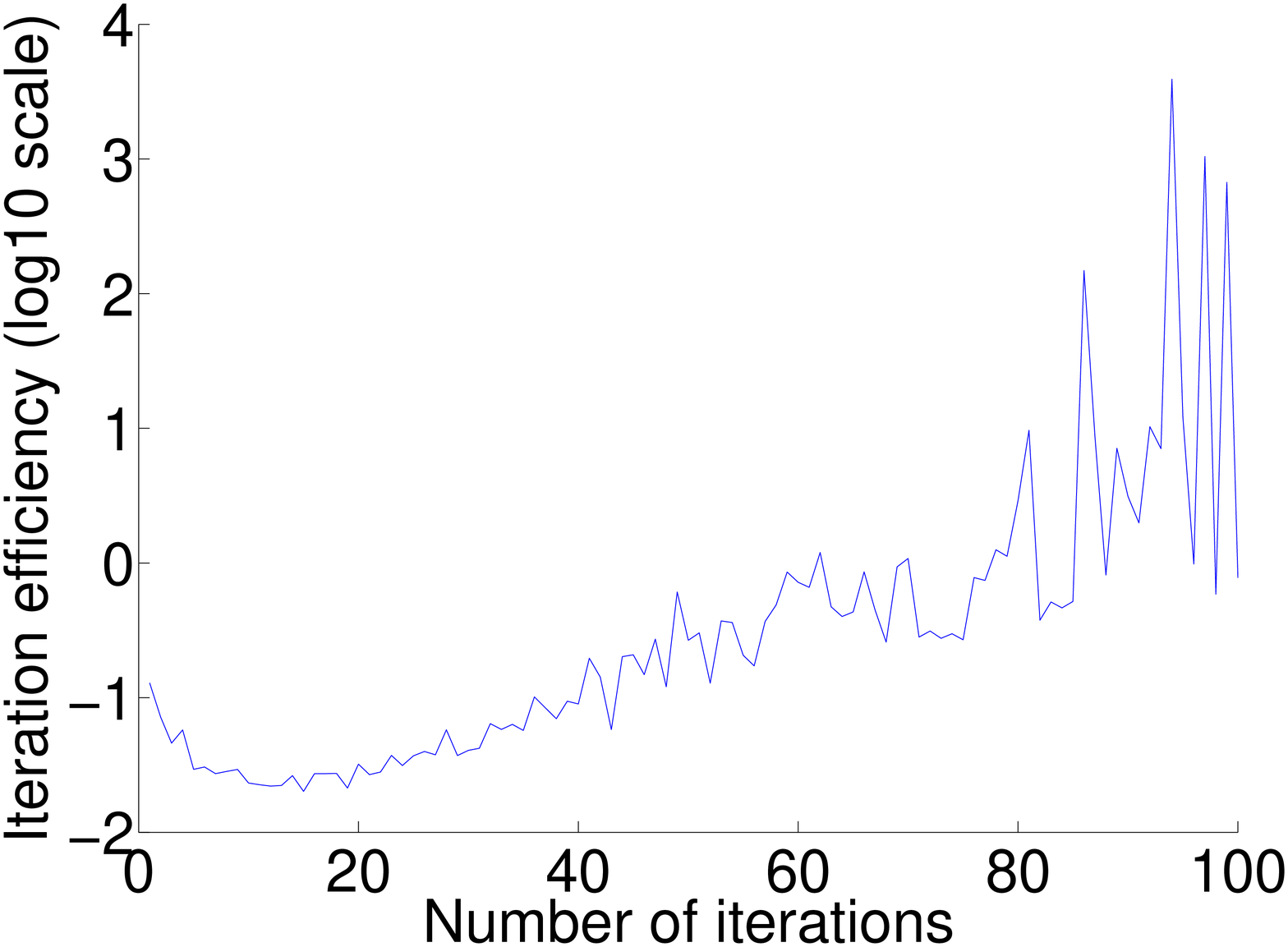}
\end{tabular}
}
{The estimated probability of convexity, the iteration time (in seconds), and the log (base 10) efficiency (left to right) of the vanilla Monte Carlo, change of measure, and conditional Monte Carlo methods (top to bottom) applied to a 30-dimensional strictly convex function. \label{fig:convex_30d_cm}}
{}
\end{figure}

Consider now the 30-dimensional test function $f(\xv) = ||\xv||^{2}, \xv \in [-10, 10]^{30}$ with $r = 3(d + 1) = 93$ sample points. The covariance matrix $\Gamma$ has diagonal entries $\Gamma_{ii} = 0.04 f^2(\xv_i)$, and off-diagonal entries $\Gamma_{ij} = 10^{-2}\exp\{-||\xv_i - \xv_j||^2 / 2\} 0.04f(\xv_i)f(\xv_j)$. Hence the variance depends on the function value, and there is also modest positive correlation between any two design points depending on the distance between them. As before, for the change of measure method, a new set of samples is obtained every iteration for the first 30 iterations, and every 5 iterations thereafter, and for the acceptance-rejection method we start to reuse samples only after the first 30 iterations. We find that the change of measure and acceptance-rejection methods do not work very well on this example. Indeed, according to Proposition \ref{thm:LRHeavyTail}, the heavy-tail behavior of the likelihood ratio becomes more severe with more design points. With the likelihood ratio often taking very large values, the change of measure estimates evaluate to large values with wide confidence intervals, as shown in Figure \ref{fig:convex_30d_cm} (notice the y-axis scale). Due to the same reason, the acceptance-rejection method reduces to vanilla Monte Carlo by rejecting almost all previous samples, so we omit that method from the results in Figure~\ref{fig:convex_30d_cm}. In early iterations, conditional Monte Carlo takes more than 6 minutes to generate an estimate using {\tt CVX} with {\tt Gurobi} ({\tt linprog} takes over 1 hour), and the iteration efficiency is around 0.20. In comparison, the vanilla Monte Carlo method only takes 80 seconds per iteration at the beginning of the iteration by solving the decomposed \ref{LS:infeasDecomp}, giving around the same level of iteration efficiency. However, towards the end of the 100 iterations, conditional Monte Carlo is able to reduce the variance of the estimated probability so well that the efficiency improves beyond that of vanilla Monte Carlo. Therefore we recommend using conditional Monte Carlo (with {\tt CVX} + {\tt Gurobi}) if one can afford the running time, and vanilla Monte Carlo otherwise or when {\tt CVX} is not installed.

\subsection{A Non-Convex Function}

\begin{figure}[H]
\FIGURE
{
\centering
\begin{tabular}{c c c}
\includegraphics[width=54mm]{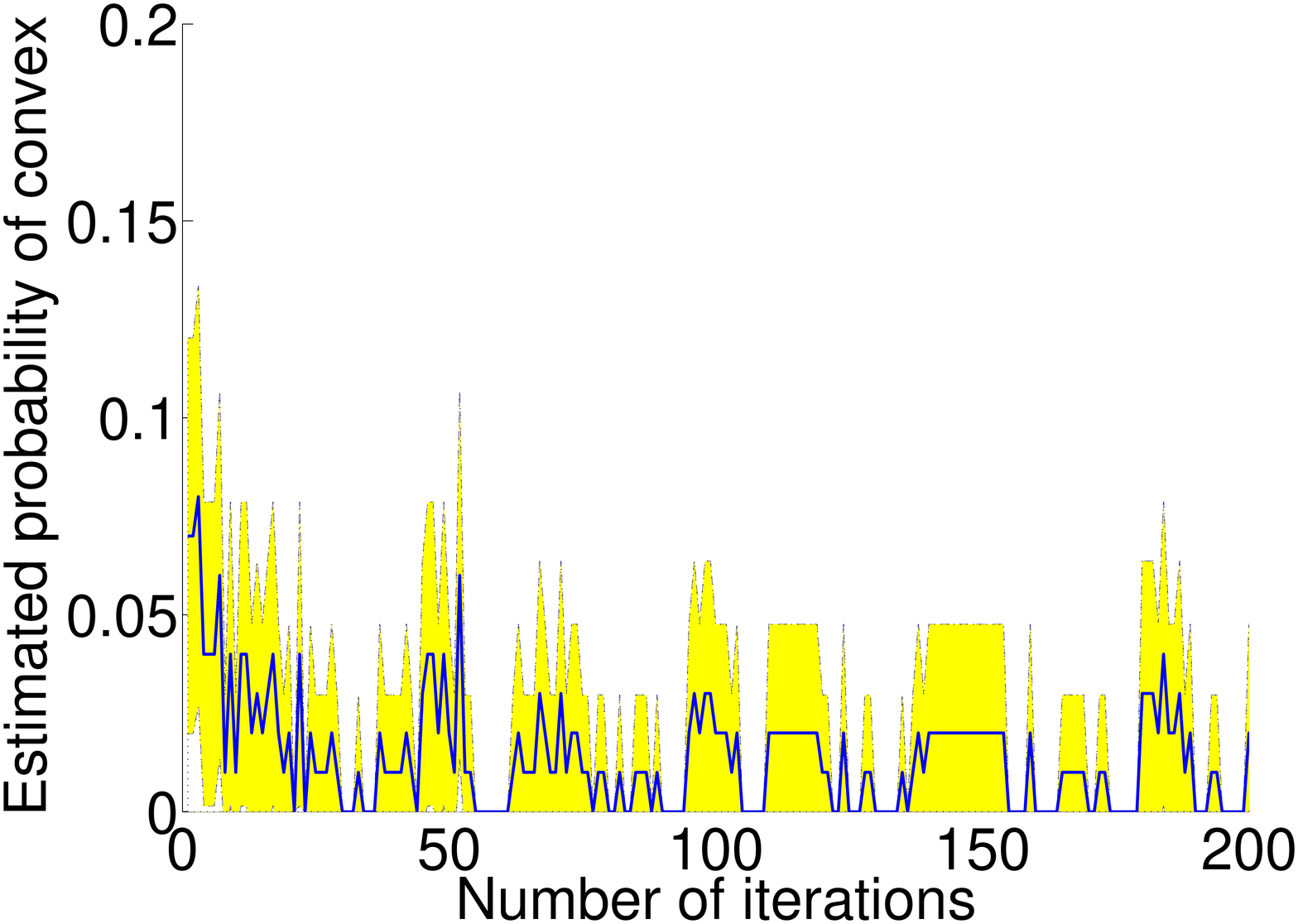} &
\includegraphics[width=54mm]{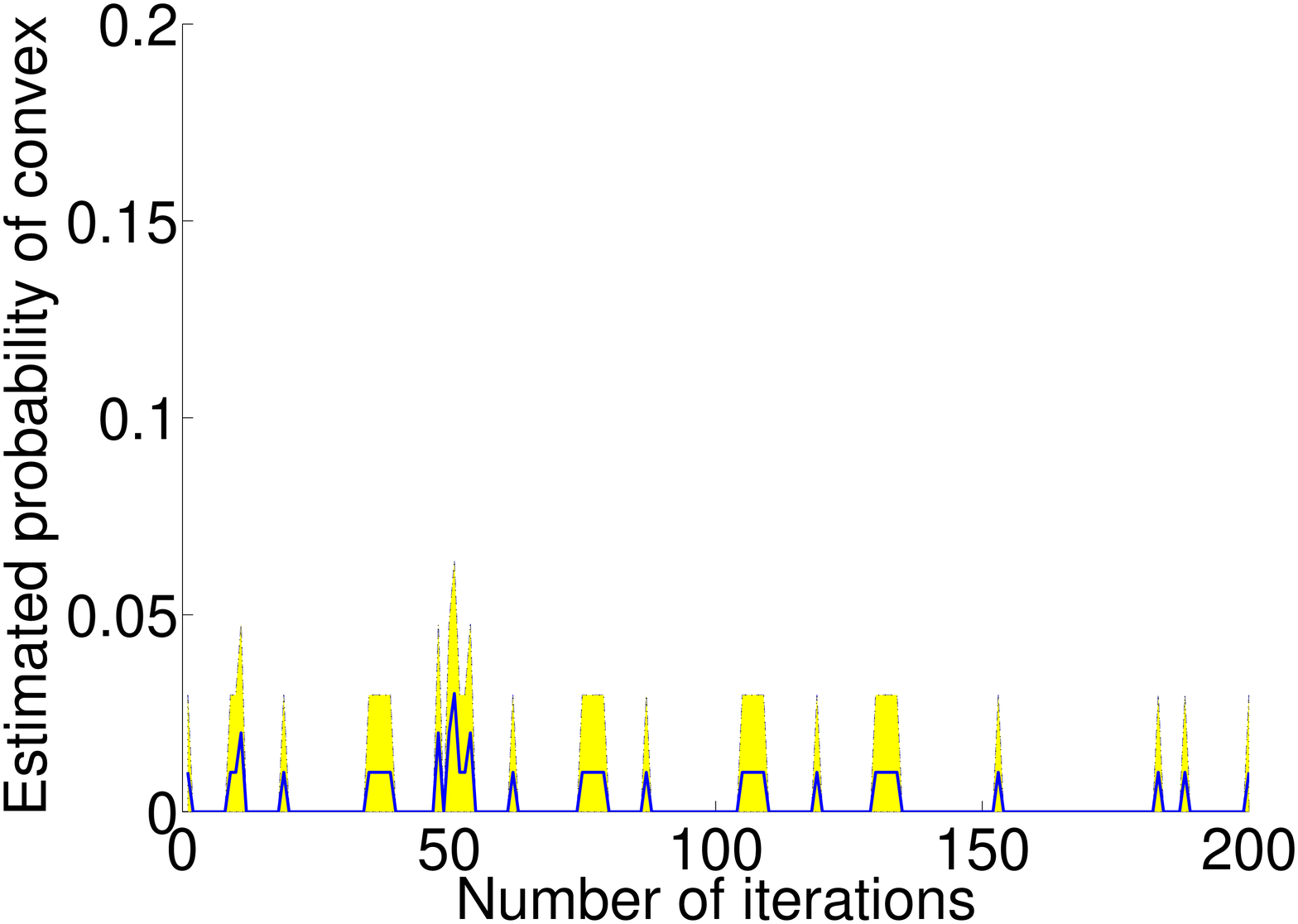} &
\includegraphics[width=54mm]{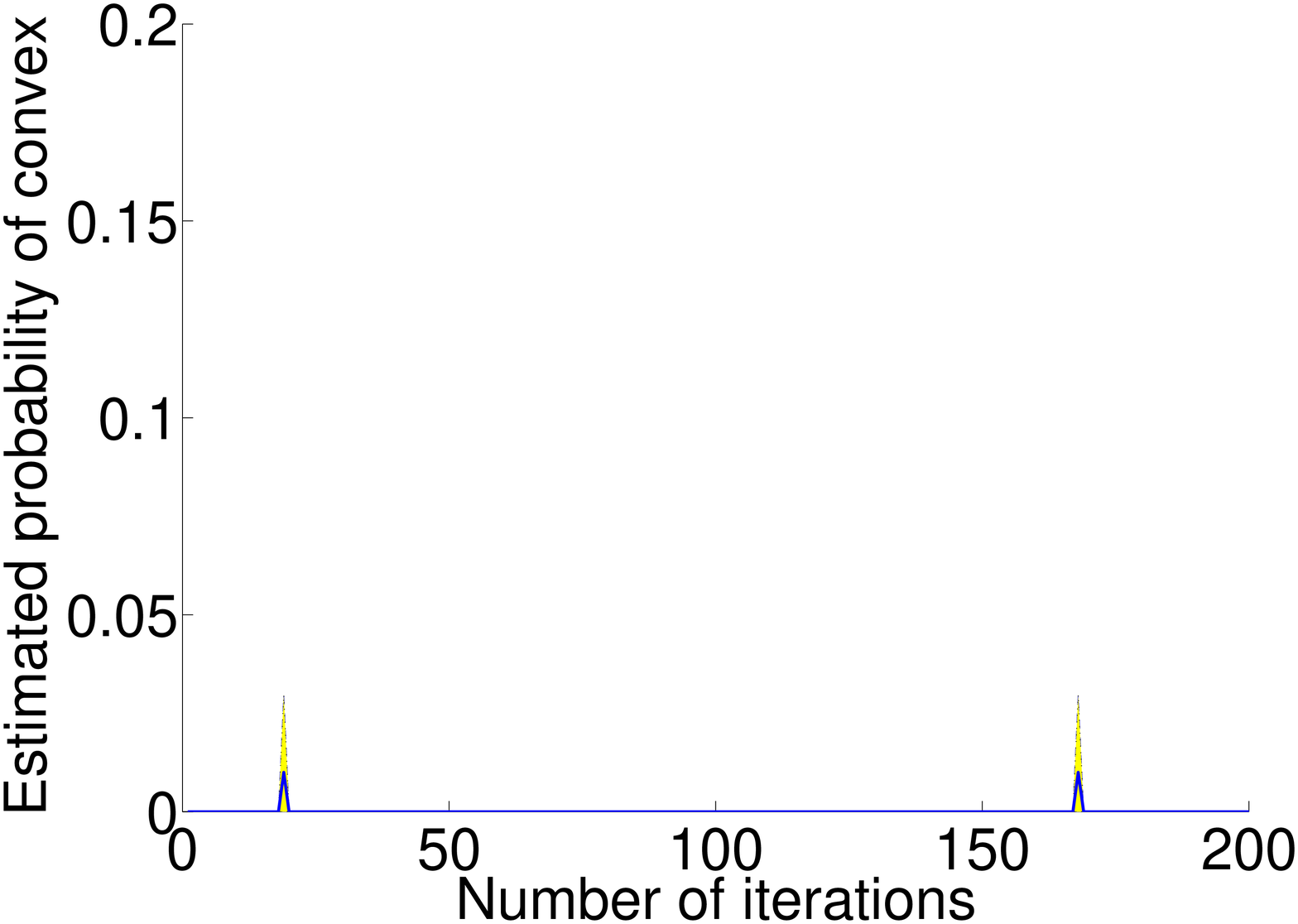}
\end{tabular}
}
{The estimated probability of convexity for the simple strictly non-convex function in dimensions 3 and 5 and 10 (from left to right). The estimates when the function is one-dimensional are not shown because they were effectively identically 0. \label{fig:nonconvex}}
{}
\end{figure}

Consider the function $f(\xv) = - ||\xv||^2, \xv \in [-1,1]^d$. In order to make the problem ``harder,'' we choose the covariance matrix $\Gamma$ to be $d^2/4$ on the diagonal, so that the sampling standard deviation is bigger than half of the function value, and $0$ on the off-diagonal. Figure~\ref{fig:nonconvex} gives the results from the vanilla Monte Carlo estimator, with acceptance-rejection applied after the initial 30 iterations, for varying dimensions. The estimated probabilities of convexity hover near zero over all iterations, especially in lower dimensions. This is perhaps intuitive: with few iterations the noise in the estimated function values dominates, and in the presence of large noise {\em any} function will appear to be nonconvex, while after many iterations, the nonconvexities of the (true) function dominate and are detected.

\subsection{Linear Function}
Linear functions are convex but lie on the boundary of the cone $\mbC$. Theorem \ref{thm:converges} does not inform us of the likely behavior of our algorithm in this case, because the event that the function lies on the boundary of $\mbC$ has measure 0 in the context of that result. Thus the posterior probability of convexity could converge to any number between 0 and 1 or not converge at all. Here we use a one-dimensional linear function $f(\xv) = 0, \xv \in [-1,1]$, with a sampling covariance matrix that equals $10^{-4}$ on the diagonal and 0 on the off-diagonal.

\begin{figure}[H]
\FIGURE
{\includegraphics[scale=0.25]{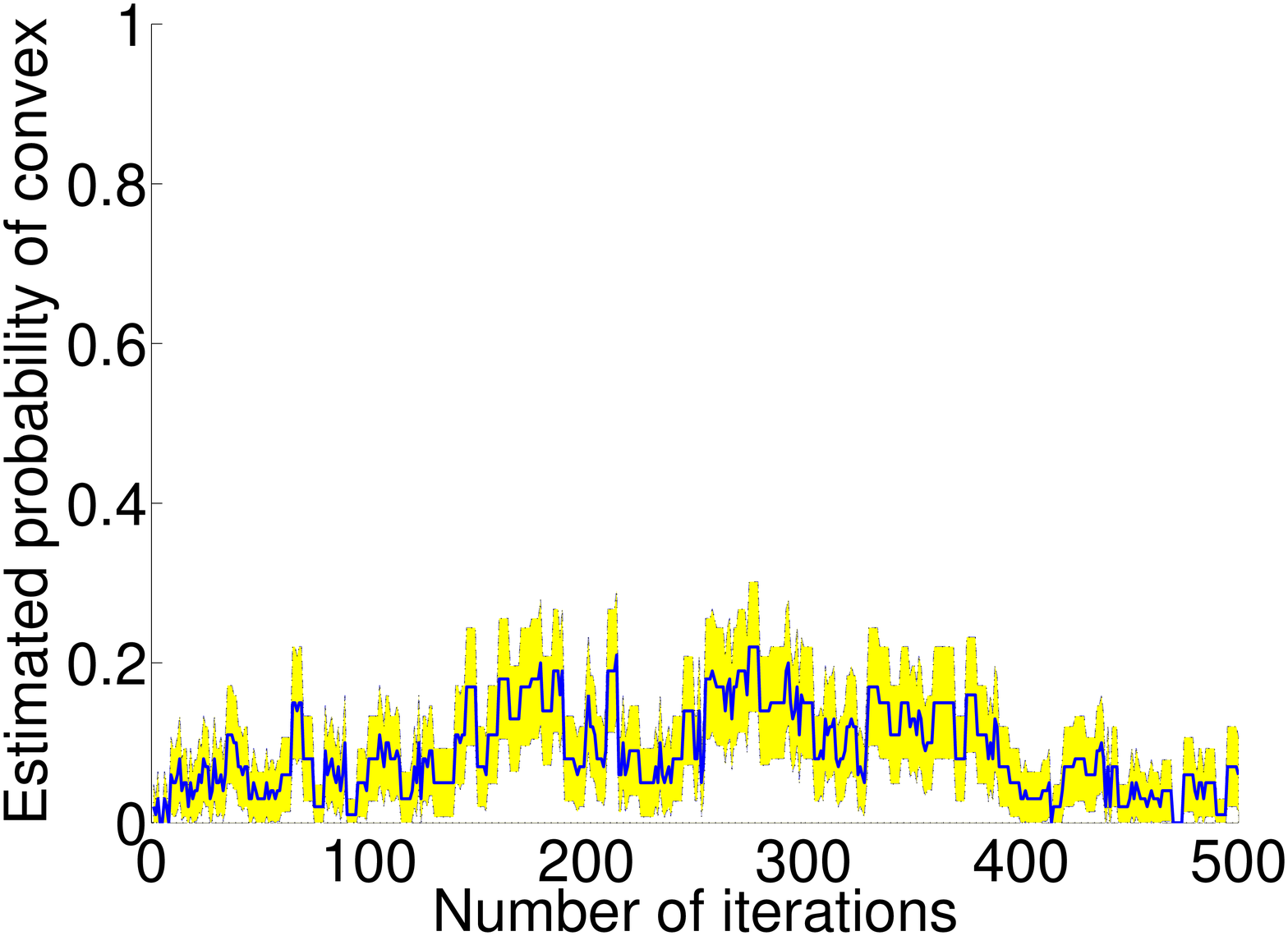}}
{The estimated probability of convexity for a 1-dimensional linear function. The mean does not appear to converge. \label{fig:linear}}
{}
\end{figure}

As shown in Figure \ref{fig:linear}, the estimated probability does not converge to 0 or 1, but stays close to 0. When we increase the dimension, e.g., to 5, and keep the sampling variance the same, the estimated probabilities stay at 0 throughout the first 100 iterations. Changing the sampling variance does not change the qualitative nature of results because when the function is zero-valued the sampling variance only changes the ``scale'' of the observations. These results are perhaps to be expected because a linear function would only appear convex when the function noise at all design points ``happens to'' form a strictly convex function. 

\subsection{Output of a Simulation}
Finally, we have also tested our algorithm on a more realistic example similar to the ``Ambulances in a Square'' problem from SimOpt \citep{ambulanceBases}. In this problem, patient calls arrive in a one kilometer unit square $[0,1]^2$ according to a Poisson process at a constant rate of 1 call every 2 hours. The $(x, y)$ locations of the calls are i.i.d.\ and distributed with a density proportional to $1.6 - (|x - 0.8| + |y - 0.8|)$. Upon receiving a call, a nearest ambulance is dispatched, traveling to the scene at a constant speed of 60 km/h. Once arriving at the scene, the ambulance spends a Gamma-distributed scene time with mean 45 minutes and standard deviation 15 minutes, then returns to the base at a speed of 40 km/h if no other call is received. We are interested in the mean response time (time from when the call is received until the ambulance arrives at the call location) as a function of the location(s) of the ambulance base(s).

\begin{figure}[htb]
\FIGURE
{
\centering
\begin{tabular}{c c c}
\includegraphics[width=54mm]{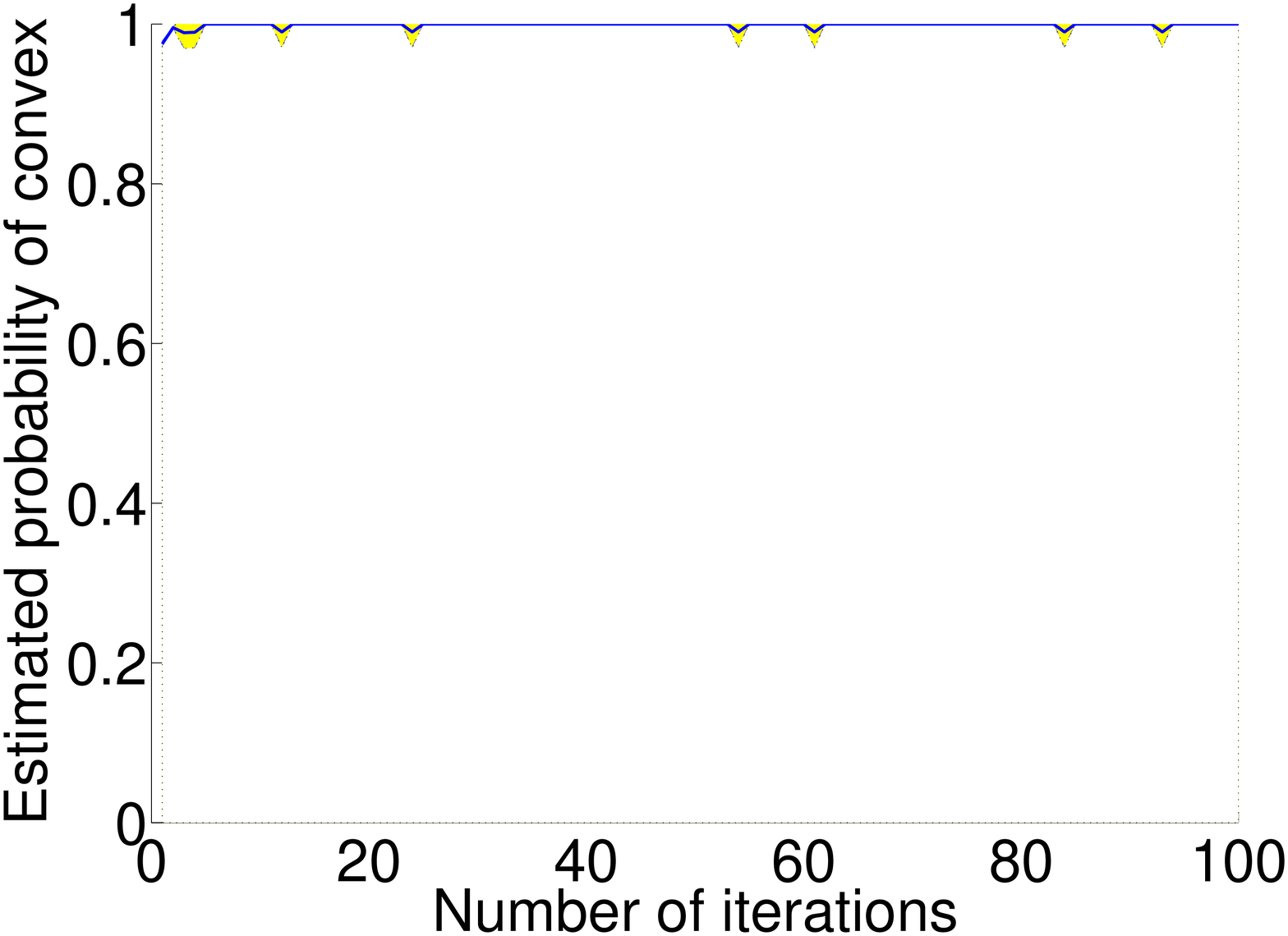} &
\includegraphics[width=54mm]{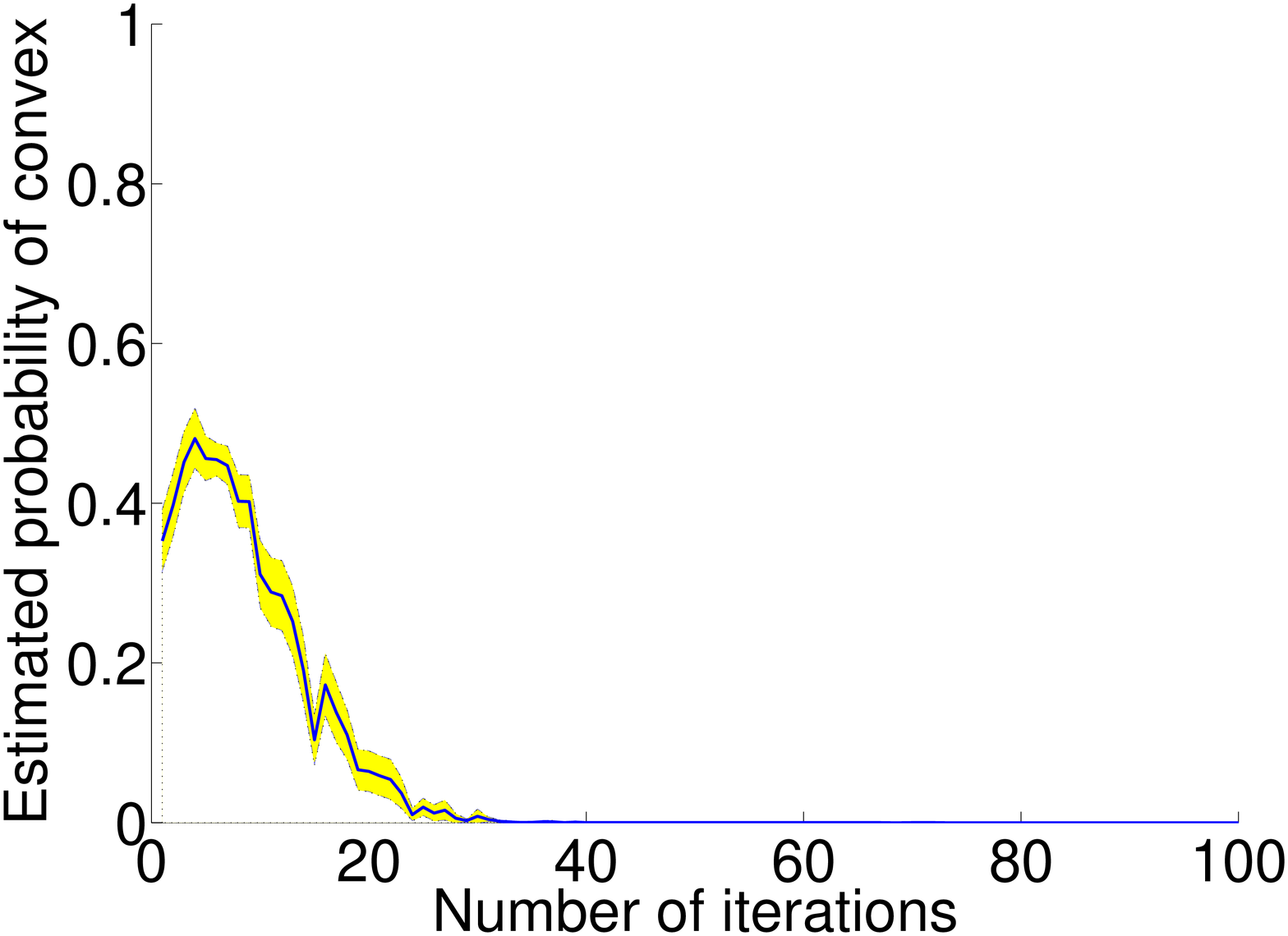} &
\includegraphics[width=54mm]{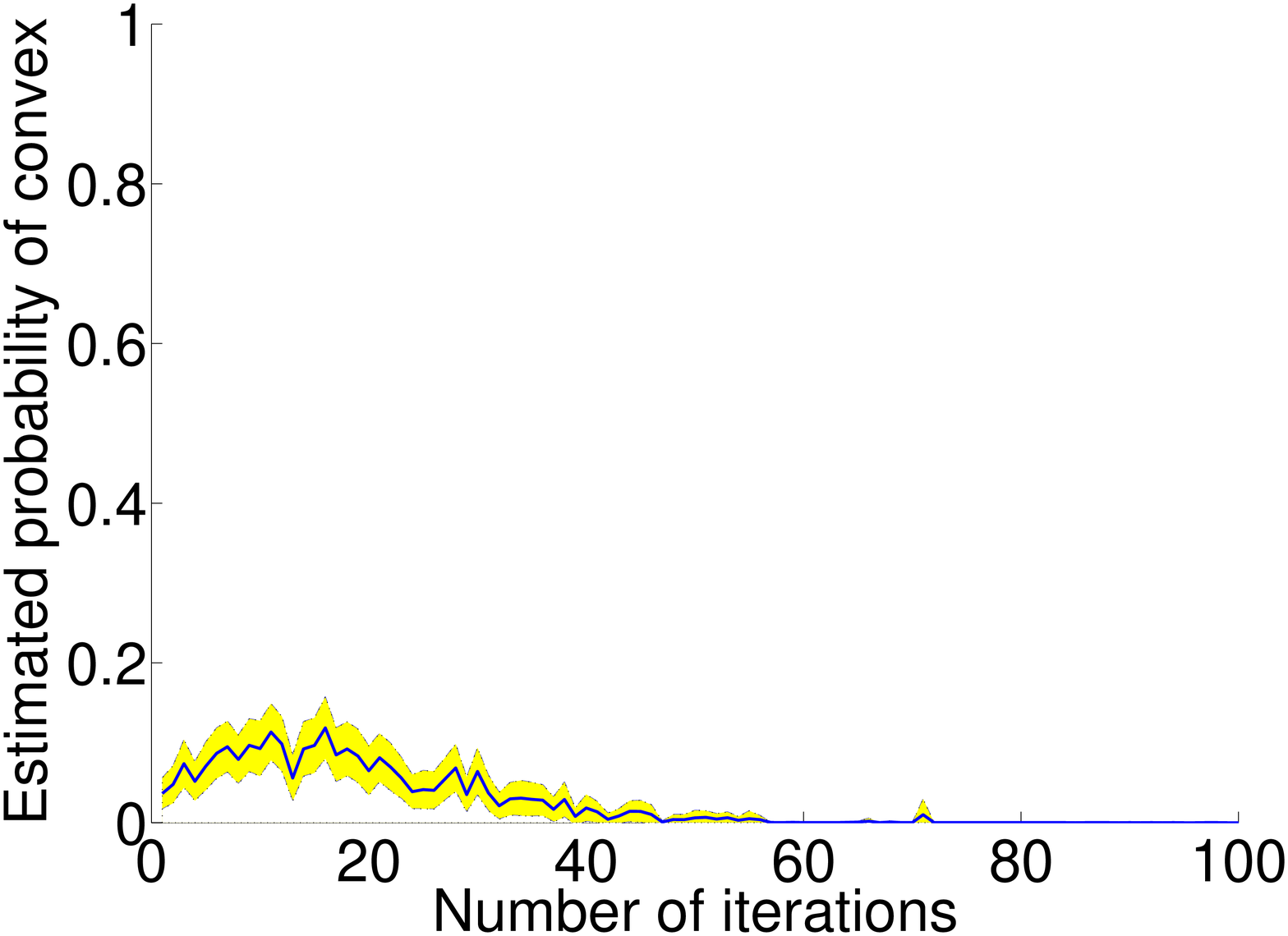}
\end{tabular}
}
{The estimated probability of convexity for the mean response time as a function of the base locations, when the number of bases is one, two, and three. \label{fig:ambulance}}
{}
\end{figure}

We sampled the base locations of the ambulance along (4 $\times$ the number of bases + 1) random lines in the unit square, with 3 points sampled on each line. Each base has two coordinates, so this is equivalent to $3(2d+1)$ design points, where $d$ is the dimension of the sample space. We are using more design points than in our previous test cases because we wanted to try more points (and consequently more computation) on a real case. Similar to our other experiments, we obtain a sample of the mean response time on each set of sampled base locations from running the simulation until 360 calls receive a response (approximately 30 days). The mean response times of the sampled base locations are evaluated using common random numbers, which compares the locations using the exact same random call arrivals and scene times. The convexity of the mean response time as a function of the ambulance base locations is tested with one, two, and three ambulance bases, using the conditional Monte Carlo method. The estimated probabilities vs.\ iteration are plotted in Figure~\ref{fig:ambulance}.

It seems that the mean response time is convex as a function of the base location when there is only one base, while it is not convex for more than one ambulance base. This agrees with our intuition that the location of a single ambulance base should have one global minimizer in the unit square. By plotting the posterior mean function, we found that the minimizer is located near the point $[0.46, 0.54]$, near, but slightly offset from, the mode $[0.8, 0.8]$, to balance the travel time to the farther corner $[0,0]$. However, when there is more than one ambulance base, the objective does not have a single minimizer due to symmetry and the interactions between bases.

\section{Conclusion}
\label{sec:conclusion}
Given a function that can be observed on a finite number of points in the presence of noise, we have suggested a sequential algorithm to estimate the posterior probability that the function is convex. The method models the function values on a fixed set of design points using a Bayesian conjugate model, and estimates the probability of convexity by Monte Carlo simulation, using samples of the function values from the posterior distribution. This Bayesian procedure gives sequential estimates for function convexity. It is useful when a function is expensive to evaluate, e.g., the output of a large simulation, or when its values can only be obtained on a constrained set of points, e.g., a function defined on a discrete domain, and is primarily an exploratory tool to help an analyst develop an understanding of the geometry of an optimization problem.

To improve the efficiency of our algorithm we introduced three variance reduction methods - change of measure, acceptance-rejection, and conditional Monte Carlo. The first two methods reuse samples obtained in an earlier iteration to calculate an estimator in the current iteration. However, they both rely on the likelihood ratio of normal or Student-$t$ posterior densities, which we prove could take extreme values due to its heavy-tail behavior. In our computational results, we observe that the change of measure method may give poor (e.g., greater than 1) estimates of the probability, and the acceptance-rejection method rejects most of the earlier samples and reduces to vanilla Monte Carlo when the number of design points is large. Finally, the conditional Monte Carlo method takes the longest time to compute but is the most effective in variance reduction, giving the highest efficiency among all methods. We recommend using it with {\tt CVX} and {\tt Gurobi}, especially for high-dimensional functions, to ensure reasonable computational time.

How should one choose $n$, the simulation runlength at each of the $r$ design points? In our experiments, we increased $n$ until the confidence intervals for $p_n$ appeared to remain near 1 (suggesting convexity) or 0 (non-convexity). We suggest this exploratory procedure as a reasonable rule of thumb, recognizing that it is only a heuristic. More advanced stopping rules that offer some kind of overall statistical guarantee might be possible, which might even lead to the development of an hypothesis testing procedure. However, such a goal is not in line with the exploratory aims of the present paper. Moreover, developing such rules would likely require considerable effort that we view to be beyond the scope of this paper.

In our experimentation, we also found the location of design points to be important when exploring an unknown function. Despite the fact that we only determine the convexity of a vector based on the pre-chosen design points, it would be helpful if the points are representative of the sample space $S$. Without knowing anything about the underlying function, a good starting point is to choose the design points such that they span the entire space. As we gain better knowledge with the sequential procedure, it is possible to expand the design points dynamically, and consequently have different sample sizes at each point. The method of choosing where and how much to sample in each iteration is left as an open problem.

A package containing the main algorithm and all variance reduction alternatives is available on Github \citep{Jian2017}.

\ACKNOWLEDGMENT{We thank the editorial team for very helpful comments. This work was partially supported by National Science Foundation grants CMMI-1200315 and CMMI-1537394, and Army Research Office grant W911NF-17-1-0094.%
}


\bibliographystyle{ijocv081} 
\bibliography{convexPaper} 


\end{document}